\documentclass{revtex4}
\usepackage{graphicx}
\usepackage{amsmath}
\usepackage{amsfonts}
\newcommand{\Asterisk}{{\scalebox{1.3}{\raisebox{-0.2ex}{$\ast$}}}}
\newcommand{\Lave}{\langle L \rangle}
\newcommand{\floor}[1]{\left\lfloor #1 \right\rfloor}

\begin{document}
\title{Screening like-charges in one-dimensional Coulomb
  systems: Exact results}
\author{Gabriel T\'ellez}
\affiliation{Departamento de F\'{\i}sica, Universidad de los Andes,
  Bogot\'a, Colombia}

\author{Emmanuel Trizac}
\affiliation{Laboratoire de Physique Th\'eorique et Mod\`{e}les
  Statistiques ({\it UMR CNRS 8626}),
 Universit\'e Paris-Sud, F-91405
  Orsay, France}

\begin{abstract}
The possibility that like-charges can attract each other under the mediation 
of mobile counterions is by now well documented experimentally, numerically, and analytically.
Yet, obtaining exact results is in general impossible, or restricted to some 
limiting cases. We work out here in detail a one dimensional model that retains 
the essence of the phenomena present in higher dimensional systems. 
The partition function is obtained explicitly, from which a wealth 
of relevant quantities follow, such as the effective force between the charges 
or the counterion profile in their vicinity. Isobaric and canonical ensembles are distinguished.
The case of two equal charges screened by an arbitrary number $N$ of counterions
is first studied, before the more general asymmetric situation is addressed. It is shown that
the parity of $N$ plays a key role in the long range physics.
\end{abstract}

\maketitle

\section{Introduction}

Coulombic effects are often paramount in soft matter systems, where
the large dielectric constant of the solvent (say water) invites
ionizable groups at the surface of macromolecules to dissociate
\cite{KeHP01,Levin02,Messina09}.  While a realistic treatment requires
considering three dimensional systems, interesting progress has been
achieved for lower dimensional problems where the key mechanisms can
be studied in greater analytical detail
\cite{Janco81,Forrester98,Samaj03}. In particular, a one dimensional
model was introduced in the 1960s by Lenard and Prager independently,
for which a complete thermodynamic solution was provided
\cite{Len61,Pra61,EL62}. This model has been further studied in
Ref. \cite{DHNP09}, but it turns out that some interesting features
have been overlooked in relation with the like-charge attraction
phenomenon \cite{Levin02,Varenna}. This striking non mean-field
effect, relevant for strongly coupled charged matter
\cite{Netz01,Varenna} is the thread in our study.

The paper is organized as follows. The model is first defined in section \ref{sec:equal-charges}.
It mimics the
screening of charged colloids. The Coulomb potential in one dimension
between two charges $q$ and $q'$ located along a line with coordinates
$\widetilde{x}$ and $\widetilde{x}'$ is
\begin{equation}
  v(\widetilde{x},\widetilde{x}')=-qq'|\widetilde{x}-\widetilde{x}'|
  \,.
\end{equation}
Therefore, the electric field created by one particle is of constant
magnitude. This fact simplifies the study of the equilibrium
statistical mechanics of such systems, and allows to obtain some of
its properties by simple arguments. Furthermore, it also allows for an
explicit computation of the partition function~\cite{Len61,Pra61}.
The system under scrutiny can be envisioned as a collection of
parallel charged plates, able to move along a perpendicular axis.  The
salient properties of this system can be obtained by simple arguments
which we present in section \ref{sec:equal-charges}, followed
afterwards by a more technical analysis where the explicit calculation
of the partition function is performed, first in the isobaric and then
in the canonical ensemble.  After having presented the symmetric case,
section \ref{sec:different-charges} will generalize the investigation
to the situations where the two screened charges are
different. Noteworthy is that parity of the particle number\textsc{}
considerations will play an important role in the remainder.

\section{Screening of two equal charges by counterions only}
\label{sec:equal-charges}

Consider two charges $q$ along a line located at $\widetilde{x}=0$ and
$\widetilde{x}=\widetilde{L}$. Between the charges there are $N$ counterions
of charge $e=-2q/N$ between them. Consider the equilibrium thermal
properties of this system at a temperature $T$, and as usual define
$\beta=1/(k_B T)$ with $k_B$ the Boltzmann constant. This simple model
mimics the screening and effective interaction between two charged
colloids in a counterion solution, without added salt. In one
dimension, $\beta e^2$ has dimensions of inverse length, therefore it
is convenient to use rescaled units in which all distances are
measured in units of $1/(\beta e^2)$: $x=\beta e^2 \widetilde{x}$. It is
also convenient to work with a dimensionless pressure
$P=\widetilde{P}/e^2$ where $\widetilde{P}$ is the pressure (equal to the
force, in one dimensional systems).

The potential energy (dimensionless, measured in units of $k_B T$) of
the system is
\begin{equation}
  \label{eq:pot}
  U=-\sum_{1\leq i < j \leq N} |x_i-x_j| + \left(\frac{N}{2}\right)^2 L.
\end{equation}
Before presenting the technical analysis, we start by simple and more 
quantitative considerations.

\subsection{Possibility of attraction between like-charges}
\label{sec:like-charge-attraction}

\subsubsection{A heuristic argument}
\begin{figure}
  \begin{center}
  \includegraphics[width=0.7\textwidth]{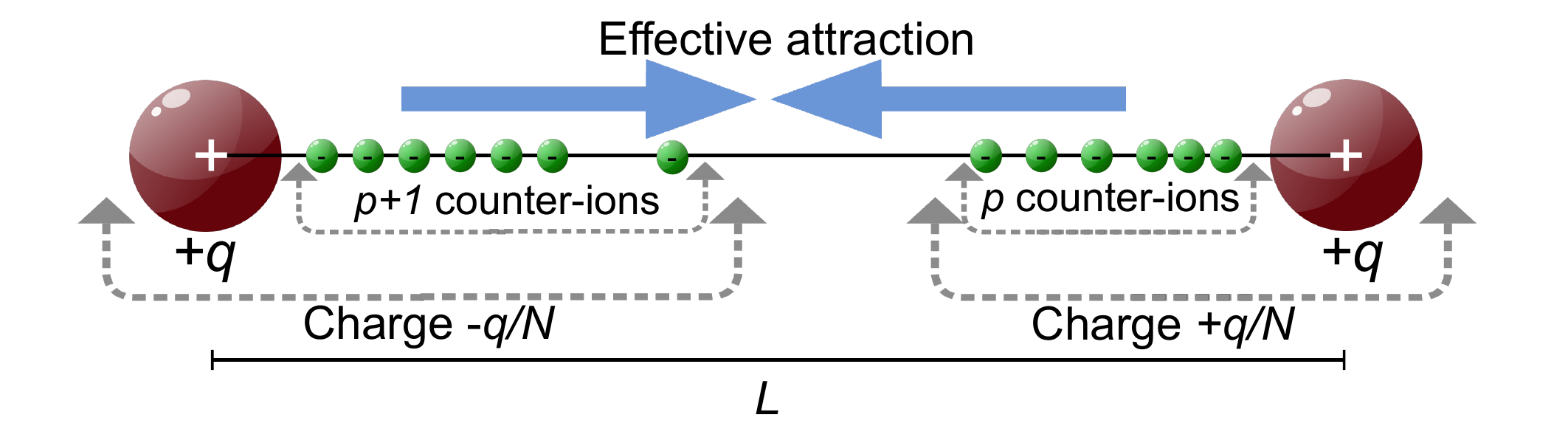}    
  \end{center}

  \caption{An odd number of mobile counterions screening two like-charges.
  The $N$ mobile ions (counter-ions) have charge $-2q/N$ and the confining objects have charge $q$,
  so that the whole system is electro-neutral. Here, $N=2p+1$ is odd, so that a single
  ion (referred to as the misfit since the net electric force acting on it vanishes) 
  ``floats'' in between the two screened boundaries which attract, each, $p$ ions in their vicinity
  (see also Fig. \ref{fig:1Dmodel-odd}).
  This single free counterion provides the binding mechanism responsible for long range attraction.
  In the canonical treatment, $L$ is held fixed, while in the isobaric situation, it is a fluctuating quantity.
    \label{fig:1Dmodel-odd-attract}}
  
\end{figure}

The possibility of attraction between the two $+q$ charges at $0$ and
$L$ is related to the parity of $N$. If $N$ is odd, $N=2p+1$, then $p$
counterions will form a double layer around each charge $q$. This
will form two compound objects with charge $q(1-2p/N)=q/N$ each one,
located around $0$ and $L$. There will be in addition one counterion between these
two object, which is essentially free, as the electric field created
by the charges located on each side around $0$ and $L$ cancel each
other. When $L$ is large enough, consider
figure~\ref{fig:1Dmodel-odd-attract}. The right side of the system
composed of one charge $q$ and $p$ counterions has charge $q/N$. The
left side which, for the sake of the argument, has the free counterion plus the compound charge, exhibits
a total charge $-q/N$. Thus the force exerted by the left side on the
right side is $\widetilde{P}\to -q^2/N^2=-e^2/4$, an attractive
force. Thus one expects that $P \to -1/4$, for $L\to\infty$.

\begin{figure}
  \begin{center}
  \includegraphics[width=0.7\textwidth]{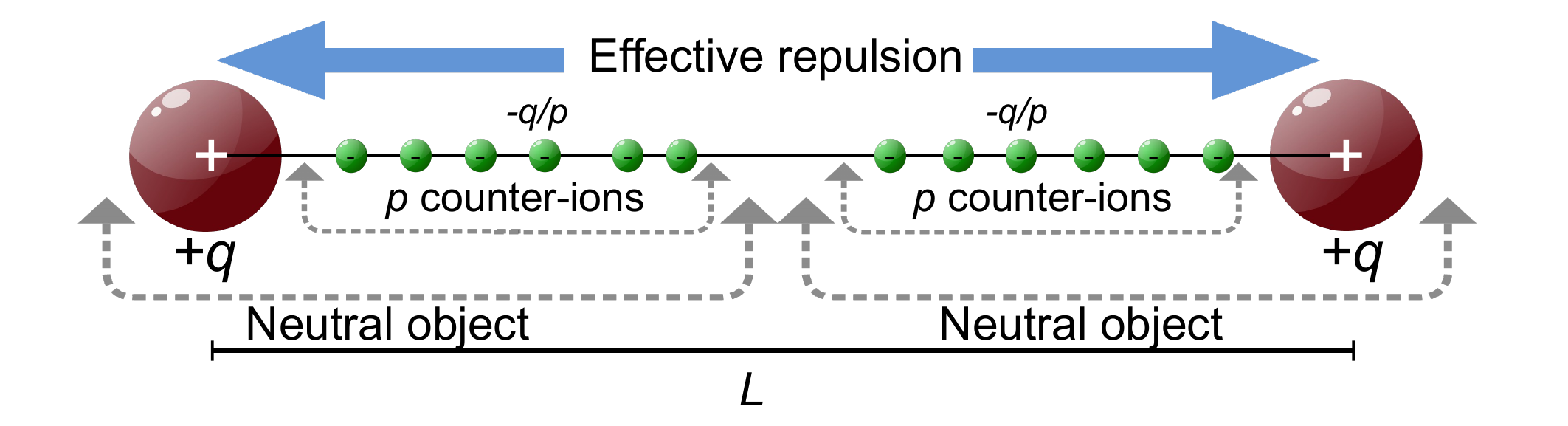}    
  \end{center}
  \caption{An even number of counterions screening two like-charges ($N=2p$).
  At large distance, the two double-layers (made up of an ion $q$ and $p$ counter-ions)
  decouple since they are neutral. No misfit ion is present to mediate attraction, and the pressure
  is repulsive at all distances.
    \label{fig:1Dmodel-even}}  
\end{figure}

On the other hand, if $N$ is even, there will not be a free counterion
between the layers, which will be completely neutral, thus one expects
that $P\to 0^{+}$ when $L\to\infty$, as shown in
figure~\ref{fig:1Dmodel-even}.

\subsubsection{Beyond heuristics}
The previous intuition, providing a large distance attraction for odd $N$,
can be substantiated by a simple calculation. Use will be made here of the contact theorem 
\cite{HBL79,contact2,contact3,DHNP09,MaTT15},
an exact relation between the force exerted on the charge $q$, and the ionic density at contact
(stemming from the mobile charges $-2q/N$). Such a relation is particularly useful
for discussing the like-charge attraction phenomenon \cite{Netz01,SaTr11,rque9}.
The argument allowing to get the contact density is two-fold, and goes as follows. 

\begin{figure}
  \begin{center}
  \includegraphics[width=0.7\textwidth]{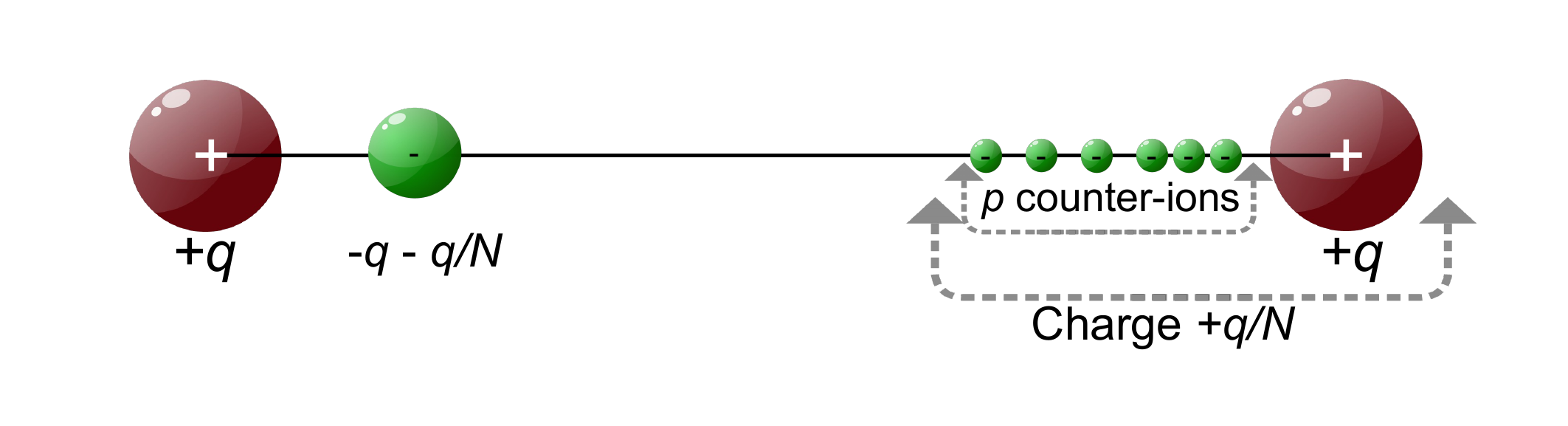}    
  \end{center}
  \caption{Upon regrouping the $p+1$ leftmost counterions in Fig. \ref{fig:1Dmodel-odd-attract}, one obtains
   an ion with charge $-q-q/N$. This newly defined system has the same large distance pressure
   as that of Fig. \ref{fig:1Dmodel-odd-attract}. 
    \label{fig:argument_trick}}  
\end{figure}

First, we argue that at large $L$, the $p$ counterions that are closest to each boundary remain in their
vicinity, while the middle free counterion (the misfit in Figs. \ref{fig:1Dmodel-odd-attract} and \ref{fig:1Dmodel-odd}), 
which does not feel any electric field by symmetry, tends to be unbounded and no longer contributes
to the pressure (discarding $1/L$ terms). In a second step, we thus compute the contact density in a system of an 
isolated charge $+q$, with a double-layer of $p$ ions in the vicinity (the total charge 
of this composite object, shown on the right hand-side of Fig. \ref{fig:1Dmodel-odd-attract}) is $q/N$. 
The solution to this problem is not immediate, but can be found by a convenient mapping
onto a more convenient problem, shown in Fig. \ref{fig:argument_trick}. As illustrated in the figure,
we regroup the $p+1$ leftmost counterions in a single ion, having charge $-q(1+1/N)$. 
At large distances, this regroupment does not influence the distribution of counterions around
the rightmost ion $+q$, and thus leaves the large $L$ pressure unaffected. The next important argument is that the pressure can be
equivalently computed from the contact density at the rightmost, or leftmost charge $+q$. 
It is thus simpler to perform the calculation in the newly defined regrouped system (left hand side of Fig.
\ref{fig:argument_trick}). 
The regrouped ion with charge $-q(1+1/N)$ is in the electric field of the charge 
$q$ on its left, and of the composite system on its right having charge $q/N$. 
This amounts to a field $q(1-1/N)$. 
Hence, the electric potential energy reads $q^2(1-1/N) \tilde{x}(1+1/N)$.
The corresponding Boltzmann weight gives the density of the regrouped ion
\begin{equation}
\rho(\tilde{x}) \,=\, \beta q^2\left(1-\frac{1}{N^2}\right) \, 
\exp\left[-\beta q^2 \tilde{x} \left(1-\frac{1}{N^2}\right)
\right]
\end{equation}
where due account was taken of normalization ($\int
\rho\,d\tilde{x}=1$). The contact density $\rho(0)=\beta q^2(1-1/N^2)$
finally yields the pressure through the contact theorem $\beta
\widetilde P = \rho(0) - \beta q^2$.  We get here $\widetilde
P=-q^2/N^2$ (or equivalently $P=-1/4$), a result which by construction
holds in the large $L$ limit.  The reason for a non vanishing pressure
at large distance is that the $p$ counter-ions cannot exactly screen
the charge of an ion $q$.  It is no longer the case when $N$ is even,
in which case $P\to 0$ for $L\to\infty$. The present results will be
fully corroborated by direct partition function calculations.

\subsubsection{Correction to large distance asymptotics and crossover pressure} 
\begin{figure}
  \begin{center}
  \includegraphics[width=0.7\textwidth]{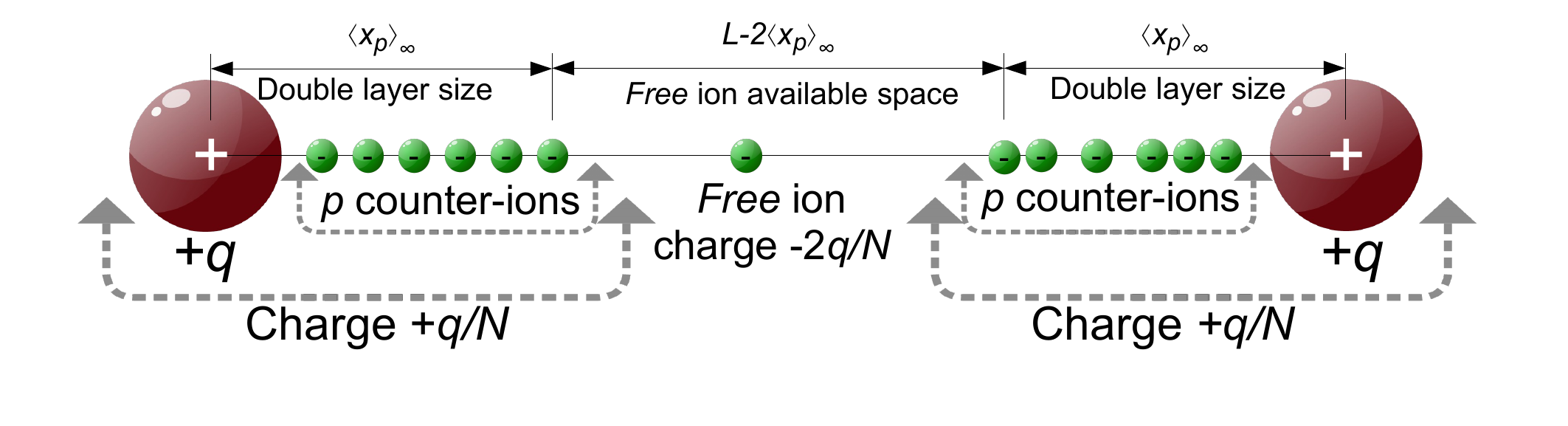}    
  \end{center}

  \caption{An odd number of counterions screening two like-charges.
    The free ``misfit'' ion is singled out.
    \label{fig:1Dmodel-odd}}
  
\end{figure}

Returning to the case when $N=2p+1$ is odd, we can also estimate the
first correction to the pressure for large $L$. Consider that $L$ is
fixed (canonical ensemble) and large. Since the system is somehow
equivalent to two double layers with a free counterion in between,
this counterion will contribute to the pressure (denoted as $P_c$ in
the canonical, fixed-$L$ ensemble) with a correction $1/L$. This estimate can be made more
quantitative. The available space for the free
counterion is not $L$, but it is rather $L$ minus the space occupied
by the diffuse counterion layers, given by $\langle x_p
\rangle_{\infty}$ the thermal average position of the $p$-th
counterion if they have been ordered $x_1<x_2< \cdots < x_p < x_{p+1}
< \cdots < x_{2p+1}$, in the limit $L\to\infty$. Thus
\begin{equation}
  P_c = -\frac{1}{4} + \frac{1}{L-2\langle x_p \rangle_{\infty}}
  + o\left(\frac{1}{L}\right)
  \,.
\end{equation}
This is illustrated in figure~\ref{fig:1Dmodel-odd}. In the following
section, we evaluate explicitly $\langle x_p \rangle_{\infty}$ and
find
 \begin{equation}
   \label{eq:xp}
   \langle x_p \rangle_{\infty} = \frac{p}{p+1} = 
   \frac{N-1}{N+1}
   \,.
 \end{equation}
Then, for large $L$, we expect
\begin{equation}
  \label{eq:pLinfty}
  P_c = -\frac{1}{4} + \frac{1}{L-2\,
    \frac{N-1}{N+1}}+ o\left(\frac{1}{L}\right)
  \,.
\end{equation}

In the other limiting case $L\to 0$, the result is \cite{DHNP09}
$P_c=N/L$, that can be understood as all the $N$ counterions are
squeezed in a small distance $L$. Thus we see that the pressure is
positive (repulsive force) for small separations $L\to 0$ then changes
to negative pressure (attractive force) for large $L$.

\begin{figure}
\begin{center}
  \includegraphics[width=0.5\textwidth]{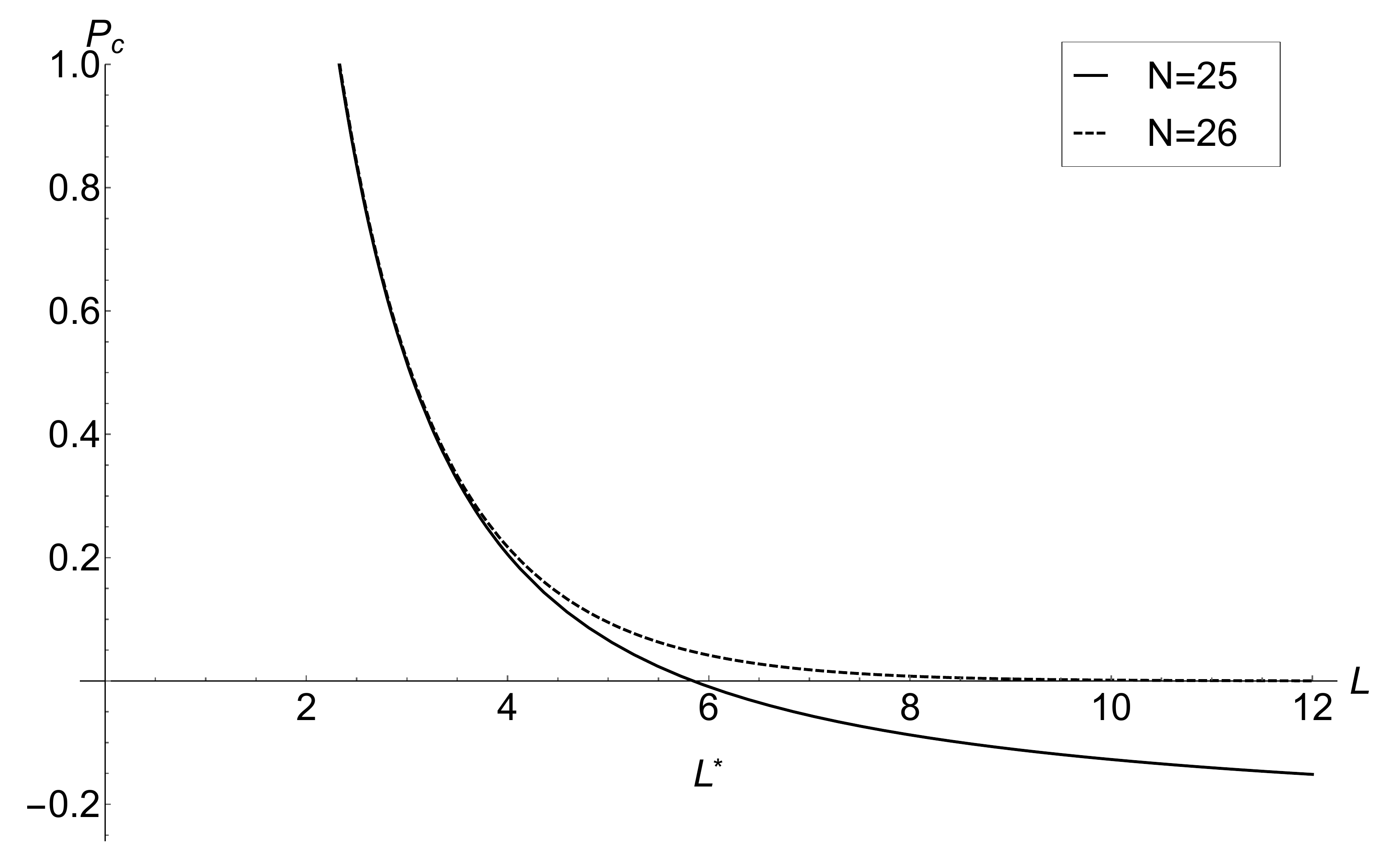}
  \caption{
    \label{fig:PvsL-canon}
    The (canonical) pressure $P_c$ as a function of the separation
    $L$, for $N=25$ (continuous bottom line) and $N=26$ (dashed upper
    line). For $N$ odd the pressure becomes negative at large
    distances. }
\end{center}
\end{figure}

We will show in the following section that the $o(1/L)$ corrections in
(\ref{eq:pLinfty}) are actually exponentially small, in the canonical
ensemble, therefore equation~(\ref{eq:pLinfty}) gives a fairly good
approximation for the pressure for a large set of values of the
separation $L$. From this, one can estimate the distance $L^*$, at
which the effective force between the two charges becomes attractive
\begin{equation}
  \label{eq:Lstar_cano}
  L^* \simeq 4 + 2   \langle x_p \rangle_{\infty} 
  = 4+ 2\,\frac{N-1}{N+1}
  \,. 
\end{equation}
Figure~\ref{fig:PvsL-canon} shows the pressure $P_c$ as a function of
$L$, for $N=25$ and for $N=26$ particles. For $N=25$ (odd) the pressure
changes its sign at $L^*=4+2*24/26\simeq 5.85$, while for $N=26$ the
pressure is always positive.

Summarizing, in the case of odd $N$, the possibility of having an
effective attraction for large separations $L$ is due to the sharing
of the ``free'' ion which leads to the creation of opposite charges
objects (ions $q$ plus their counterion clouds). Although the
analytical results presented here are valid only for this
one-dimensional model, the same physical mechanism has also been
observed in three dimensional systems~\cite{MHK00,Kim14}.
It can also be surmised that in situation of odd $N$ where the
free counterion has a varying charge, attraction will be all the stronger
as the charge will increase in absolute value. In addition, the very mechanism
brought to the fore here indicates that at mean-field level, where the
discrete nature of ions is discarded, attraction should be suppressed,
which indeed is the case \cite{lca1,lca2,lca3}.

%%%%
%
%
%%%%%

\subsection{Explicit exact calculation of the partition function}

\subsubsection{Preliminary observations}

The equilibrium thermodynamics of the one-dimensional two-component
Coulomb gas was solved simultaneously but independently by
Lenard~\cite{Len61} and Prager~\cite{Pra61}. In the present model,
only one type of identical particles (the counterions) are present. It
is convenient to order the particles as $0\leq x_1 \leq \cdots \leq x_N
\leq L$. Then, rearranging the terms in (\ref{eq:pot}), the potential
energy of the system can be written as
\begin{equation}
  \label{eq:Uodd}
  U=\frac{N^2 L}{4}-2\sum_{j=0}^{p-1}(p-j)(x_{2p+1-j}-x_{1+j})
\quad
\text{for\  }N=2p+1\text{\ odd},
\end{equation}
and
\begin{equation}
  U=\frac{N^2 L}{4}-\sum_{j=0}^{p-1}(2p-2j-1)(x_{2p-j}-x_{1+j})
\quad
\text{for\  }N=2p\text{\ even}.
\end{equation}
Notice that in the case $N=2p+1$, the particle with position $x_{p+1}$
does not appear in the potential energy. It is the free counterion (misfit)
discussed in the previous section, whose role is crucial for the
possibility of like-charge attraction.

The canonical configuration integral is
\begin{equation}
  \label{eq:Z}
  Z_c(N,L)=\int_{0}^{L} dx_N \int_0^{x_N} dx_{N-1}\cdots
  \int_{0}^{x_3} dx_2 \int_{0}^{x_2} dx_1 \,
  e^{-U}
  \,.
\end{equation}
As mentioned by Lenard in his seminal paper~\cite{Len61} ``the
(configuration) integral is elementary (because) the class of
functions consisting of exponential of linear functions is closed
under the operation of indefinite integral (...) however the task of
evaluating (it) is not trivial''. For small $N$ one can compute by
hand $Z_c$, and for larger given values of $N$ it can be obtained
numerically with the aid of a computer algebra system software program.  By
inspection of the integral (\ref{eq:Z}), one can deduce that $Z_c$ is
a linear combination of products of exponentials of $L$ and linear
functions of $L$. One can also deduce the argument of each
exponential function of $L$ by keeping track of the factor that
multiplies each $x_k$ in the integral (\ref{eq:Z}).  These come from
the explicit term in $U$ (for instance, for $x_{j+1}$ it is
$2(p-j)$ in the case $N$ odd), but after each successive
integration, the factor of $x_{k}$ will by added to the one of
$x_{k+1}$ due to the upper limit of integration. Taking that into
account, one realizes that the exponentials of $L$ in $Z_c$ are of the
form $\exp(-(j+\frac{1}{2})^2 L)$ in the case $N$ odd, and $\exp(-j^2
L)$ in the case $N$ even. Thus, the configuration canonical
integral is expected to be of the form
\begin{equation}
  Z_c(N,L)=\sum_{j=0}^{p}  e^{-(j+\frac{1}{2})^2 L } ( A_j L + B_j)
\quad
\text{for\  }N=2p+1\text{\ odd},
\end{equation}
and 
\begin{equation}
  Z_c(N,L)=\sum_{j=0}^{p}  e^{-j^2 L } ( C_j L + D_j)
\quad
\text{for\  }N=2p\text{\ even}.
\end{equation}
The non trivial task is to evaluate explicitly the coefficients $A_j$,
$B_j$, $C_j$ and $D_j$. This is done in
section~\ref{sec:canonical-exact}.

\subsubsection{Previous results}

In~\cite{DHNP09}, the present system was studied, but an exact analytical
explicit evaluation of the partition function for an arbitrary number
of particles was not achieved. Rather, an interesting reformulation of
this model was proposed, by mapping it into a quantum mechanical
problem, following a technique put forward by Edwards and
Lenard~\cite{EL62}. It was shown in~\cite{DHNP09} that the
configuration integral is given by
\begin{equation}
  Z_c(N,L)=b(N/2,N/2,L)
\end{equation}
where $b(n,N/2,x)$ is the solution of a set of $N$ coupled elementary
linear differential equations
\begin{equation}
  \frac{db(n,N/2,x)}{dx}=-(n^2/2)\, b(n,N/2,x)+b(n-1,N/2,x)
\end{equation}
with the initial condition $b(n,N/2,0)=\delta_{n,-N/2}$. Integrating
this equation one has
\begin{equation}
  \label{eq:Dean-system}
  b(n,N/2,x_n)=\int_0^{x_{n}} e^{-(n^2/2) (x_{n}-x_{n-1})}
  b(n-1,N/2,x_{n-1}) \, dx_{n-1}
  \,.
\end{equation}
Then, starting from the known $b(-N/2,N/2,x_1)$ one has to perform
successively $N$ integrals~(\ref{eq:Dean-system}) to obtain
$b(N/2,N/2,L)$ and the configuration integral. This task is 
equivalent to performing directly the $N$ integrals of the configuration
integral~(\ref{eq:Z}). Thus, unfortunately, the method proposed
in~\cite{DHNP09} does not provide any computational advantage over a
direct numerical evaluation of the partition function.

Here, our goal is to obtain an
explicit analytical expression for the configuration integral for an
arbitrary number of particles $N$. Using Lenard~\cite{Len61} and
Prager~\cite{Pra61} method, we will first compute the partition
function of the constant pressure ensemble
\begin{equation}
  Z_P(N,P)=\int_{0}^{\infty} e^{-P L} Z_c(N,L)\, dL
\end{equation}
which is the Laplace transform of the canonical configuration integral
$Z_c$. This is a straightforward application of the technique of
Lenard and Prager, and it is actually much simpler than the complete
work presented in~\cite{Len61,Pra61}, since all particles are
identical and we will not have to deal with the combinatorial problem
of studying the different configurations of charges.

Then, we shall invert the Laplace transform to obtain the canonical,
constant ``volume'' $L$, configuration integral $Z_c(N,L)$. Since we
are interested in finite systems, the results from the canonical
ensemble and the constant pressure ensemble will differ, and it is
of interest to compare them.

\subsubsection{Evaluation of the diffuse layer size $\langle x_p
  \rangle_{\infty}$}

To introduce the technique used to compute the partition function, we
undertake in this section a preliminary, simpler task, based on the
same technique: the exact evaluation of the diffuse layer
size $\langle x_p \rangle_{\infty}$. This quantity appeared in the
discussion of section~\ref{sec:like-charge-attraction}. Consider here
that $L\to\infty$ and $N=2p+1$. The double layer composed by the
charge $q$ at $L$ and its corresponding $p$ counterions are thereby `sent
to infinity'. The remaining $p+1$ counterions, however, still feel the
electric field created by this far charged double layer. The potential
energy part which depends on the position of the remaining counterions
is
\begin{equation}
  U_{\infty}=2\sum_{j=0}^{p-1}(p-j) x_{1+j}
  \,.
\end{equation}
We wish to evaluate
\begin{equation}
  \langle x_p \rangle_{\infty}=\frac{\int_{0<x_1<x_2<\cdots<x_p} x_p\,
    e^{-U_{\infty}} \,\prod_{k=1}^{p}dx_k}{\int_{0<x_1<x_2<\cdots<x_p}
    e^{-U_{\infty}}
    \,\prod_{k=1}^{p}dx_k}
\,.
\end{equation}
Let
\begin{equation}
  F(s)=\int_{0<{x}_1<{x}_2<\cdots<{x}_p} \, e^{-U_{\infty}- s
    x_p/2}\,d{x}_1\ldots d{x}_p
  \,.
\end{equation}
Then $\langle {x}_p \rangle_{\infty}=-2\,d\ln F(s)/ds|_{s=0}$. Following
Lenard~\cite{Len61} and Prager~\cite{Pra61} it is convenient to
re-write the potential energy as
\begin{equation}
  U_{\infty} = \frac{1}{2} \left[\sum_{j=1}^{p}
  \left( (p-j+1)^2+(p-j+2)^2 \right)(x_j-x_{j-1})
  -x_p\right]
\end{equation}
with the convention that $x_0=0$. Let us define
\begin{equation}
  \label{eq:fj}
  f_j({x})=e^{-
    \left[ (p-j+1)^2+(p-j+2)^2 \right] {x}/2 }\, H({x})
\end{equation}
where $H({x})$ is the Heaviside step function. Then
\begin{equation}
  F(s)=\int_{0}^{\infty} d{x}_1 \cdots \int_{0}^{\infty} d{x}_p
  \prod_{j=1}^{p} f_{j}({x}_{j}-{x}_{j-1}) \, e^{-(s-1){x}_p/2}
\end{equation}
We notice that $F(s)$ is the Laplace transform (evaluated at
$(s-1)/2$) of the $p$-fold convolution product $f_1*f_2*\cdots *
f_p$. The Laplace transform ${\cal L} f_{j}$ of $f_j$ is elementary
\begin{equation}
  {\cal L} f_{j}\left(\frac{s-1}{2}\right)
  =\frac{2}{(p-j+1)^2+(p-j+2)^2+s-1}=
  \frac{2}{2(p-j+1)(p-j+2)+s}
\end{equation}
Then
\begin{equation}
  F(s)=\prod_{j=1}^p\frac{2}{2(p-j+1)(p-j+2)+s}
  =\prod_{k=1}^p\frac{2}{2 k (k+1)+s}
\end{equation}
Computing the derivative of $\ln F(s)$ we obtain
\begin{eqnarray}
  \langle {x}_p \rangle_{\infty}&=&
-2\left.\frac{d\ln F(s)}{ds}\right|_{s=0}=\sum_{j=1}^p
\frac{1}{(p-j+1)(p-j+2)}
\nonumber
\\
&=&
\sum_{k=1}^p\frac{1}{k(k+1)}=
\sum_{k=1}^p\left(\frac{1}{k}-\frac{1}{k+1}\right)
\nonumber\\
&=&\left(1-\frac{1}{p+1}\right)=
\frac{p}{p+1}
\,.
\end{eqnarray}
Thus proving (\ref{eq:xp}).

\subsubsection{Isobaric ensemble}
\label{sec:isobaric-exact}

Consider now the finite system with $L<\infty$. We will detail the
calculations in the case $N=2p+1$ odd, the case $N$ even can be
obtained by a simple adaptation of the same technique. As it was done
in the previous section, it is convenient to re-write the potential
energy (\ref{eq:Uodd}) as
\begin{equation}
  U=-\frac{L}{4}
  +\frac{1}{2}
  \sum_{j=1}^{p+1} \left( (p-j+1)^2+(p-j+2)^2 \right)
  \left(x_{2p-j+3}-x_{2p-j+2} + x_j - x_{j-1} \right)
\end{equation}
where, by convention, we defined $x_0=0$ and $x_{2p+2}=L$. 
With $f_j$ defined in (\ref{eq:fj}), we notice
again that the canonical partition function is a convolution product
of $2p+2$ functions $f_j$
\begin{equation}
  \label{eq:Zc_convol}
  Z_c(2p+1,L)= e^{L/4} \left(
  \mathop\Asterisk_{j=1}^{p+1} f_j * f_j \right) (L)
\,.
\end{equation}
The isobaric partition function $Z_P$ is the Laplace transform of
$Z_c$, and we have
\begin{eqnarray}
  Z_P(2p+1,P)&=&  
  \prod_{j=1}^{p+1} \left( {\cal L}f_j \left( P-\frac{1}{4} \right)\right)^2
  \nonumber\\
  &=&
  \prod_{k=0}^{p} \frac{4}{[2 k (k+1)+s]^2}
  =
  \prod_{k=0}^{p} \frac{1}{\left[\left(k+\frac{1}{2}\right)^2+P\right]^2}
  \label{eq:ZP_odd}
 \end{eqnarray}
 where $s=(4P+1)/2$. Factoring
 $\left(k+\frac{1}{2}\right)^2+P=(k+\frac{1}{2}-i\sqrt{P})(k+\frac{1}{2}+i\sqrt{P})=|k+\frac{1}{2}+i\sqrt{P}|^2$,
the above product can be expressed in terms of Gamma functions
\begin{equation}
  Z_P(2p+1,P)=  \left( \frac{1}{P+\frac{1}{4}} \right)^2
  \left|
    \frac{\Gamma(\frac{3}{2}+i\sqrt{P})}{\Gamma(p+\frac{3}{2}+i\sqrt{P})}
  \right|^4
  \,.
\end{equation}

The average length of the system is given by the usual
thermodynamic relation
\begin{eqnarray}
  \label{eq:Lave}
  \langle L \rangle & = & -\frac{\partial \ln Z_P}{\partial P}
  =  \frac{2}{P+ \frac{1}{4}} 
   +\sum_{k=1}^{p} \frac{2}{\left(k+\frac{1}{2}\right)^{2}+ P}
  \\
  &=& \frac{2}{P+ \frac{1}{4}} + \frac{2}{\sqrt{P}}
  \Im m\hspace{-1mm}
  \left[ \psi\left(p+\frac{3}{2}+i\sqrt{P}\right)
    -\psi\left(\frac{3}{2}+i\sqrt{P}\right)
    \hspace{-0.5mm}\right]
\end{eqnarray}
where $\psi(z)=d\ln\Gamma(z)/dz$. We can notice that this expression
has a pole for $P=-1/4$, from which we obtain the
behavior when $\langle L\rangle\to\infty$, $P\to-1/4$, in agreement
with the general discussion of
section~\ref{sec:like-charge-attraction}. When $N$ is even this pole
is absent (see below).

If $N=2p$ is even, similar calculations lead to
\begin{equation}
  \label{eq:Zc_even}
  Z_c(2p,L)=e^{L/4}
    f_{p+\frac{3}{2}} * 
      \left( \mathop\Asterisk_{j=1}^{p} f_{j+\frac{1}{2}} * f_{j+\frac{1}{2}} \right)
      (L)
\end{equation}
and
\begin{equation}
\label{eq:ZP_even}
  Z_P(2p,P)=
  \frac{1}{P}\prod_{k=1}^{p} 
  \frac{1}{(k^2+P)^2}
  =  \frac{1}{P}
  \left|\frac{\Gamma(1+i\sqrt{P})}{\Gamma(p+1+i\sqrt{P})}\right|^4
  \,.
\end{equation}
Notice an important difference in the analytic structure of the
partition function in the case $N$ odd
(\ref{eq:Zc_convol})--(\ref{eq:ZP_odd}) and $N$ even
(\ref{eq:Zc_even})--(\ref{eq:ZP_even}): for $N$ even, there is a single
function $f_{p+3/2}$ in the convolution product, leading to a pole of order
one for $P=0$, in contrast to the case $N$ odd, where the functions
$f_{p+1}$ appear twice in the convolution product and the pole for the
smallest value of $|P|$ is of order two and it is for $P=-1/4$, rather
than $P=0$. In the case $N$ even, the term $f_{p+1}*f_{p+1}$
corresponds to the coupling of the left diffuse layer with the free
counterion and the coupling of this same free counterion with the right
diffuse layer. On the other hand in the case $N$ odd, the term
$f_{p+3/2}$ corresponds to the direct coupling of the left and right
diffuse layers.

The average length, for $N=2p$ even, is
\begin{equation}
  \Lave = \frac{1}{P} +
  \sum_{k=1}^p \frac{2}{k^2 +  P}
  \,.
\end{equation}
We note that $\Lave \to \infty$ when  $P\to 0^{+}$, in contrast to what happens
when $N$ is odd, where $\Lave \to \infty$ when  $P\to -1/4$.

\subsubsection{Canonical ensemble}
\label{sec:canonical-exact}

We return to the case $N=2p+1$ odd. To compute the canonical partition
function, we need to invert the Laplace transform computed in the
previous section
\begin{equation}
  Z_c(2p+1,L)= 
  {\cal L}^{-1}\left (   \prod_{k=0}^{p}
    \frac{1}{\left[\left(k+\frac{1}{2}\right)^2+P\right]^2}
  \right) (L).
\end{equation}
This rather technical part of the analysis is presented in 
Appendix \ref{app:A}, where it is shown that 
\begin{equation}
  \label{eq:ZexactNimpar}
  Z_c(2p+1,L)=
  \sum_{j=0}^{p}
  \left[\frac{2j+1}{(p-j)!(p+j+1)!}\right]^2
  e^{-(j+\frac{1}{2})^2 L}
  \left[
    L+\frac{2}{2j+1}\left(
      \sum_{k=p-j+1}^{p+j+1} \frac{1}{k}
      -\frac{1}{2j+1}
    \right)
    \right]
  \,.
\end{equation}
From this expression, we obtain the canonical pressure $P_c=\frac{d\ln
  Z_c}{dL}$,
\begin{equation}
  \label{eq:PexactNimpar}
  P_c=-\frac{ {\displaystyle    \sum_{j=0}^{p}}
    \frac{4\left(j+\frac{1}{2}\right)^4}{\left[(p-j)!(p+j+1)!\right]^2}
    \left[
      L+\frac{2}{2j+1}\left(
      {\displaystyle \sum_{k=p-j+1}^{p+j+1} }
      \frac{1}{k}
      -\frac{3}{2j+1}
      \right)
      \right]
    e^{-(j+\frac{1}{2})^2 L}
  }{
    {\displaystyle \sum_{j=0}^{p} }
    \left[\frac{2j+1}{(p-j)!(p+j+1)!}\right]^2
    \left[
      L+\frac{2}{2j+1}\left(
      {\displaystyle \sum_{k=p-j+1}^{p+j+1}}
      \frac{1}{k}
      -\frac{1}{2j+1}
      \right)
      \right]
    e^{-(j+\frac{1}{2})^2 L}
    }
  \,.
\end{equation}

For $N=2p$ even, the results are
\begin{equation}
  Z_c(2p,L)=
    \frac{1}{(p!)^4}-
    \sum_{j=1}^p \frac{(2 j)^2
    e^{-j^2 L}}{[(p+j)!(p-j)!]^2}
    \left[
      L + \frac{1}{j}\left(
      \sum_{k=p-j+1}^{p+j} \frac{1}{k} 
      - \frac{1}{2j} 
      \right)
      \right]
  \,,
\end{equation}
and
\begin{equation}
  P_c=\frac{
    {\displaystyle\sum_{j=1}^p} \frac{4j^4
    e^{-j^2 L}}{[(p+j)!(p-j)!]^2}
    \left[
      L + \frac{1}{j}\left(
      {\displaystyle\sum_{k=p-j+1}^{p+j}}
      \frac{1}{k} -\frac{3}{2j}
      \right)
      \right]}
  {\frac{1}{(p!)^4}-
    {\displaystyle \sum_{j=1}^p } \frac{(2 j)^2
    e^{-j^2 L}}{[(p+j)!(p-j)!]^2}
    \left[
      L + \frac{1}{j}\left(
      {\displaystyle\sum_{k=p-j+1}^{p+j}} \frac{1}{k} -\frac{1}{2j}
      \right)
      \right]}
  \,.
\end{equation}

\subsubsection{Limiting cases and comparison between the ensembles}

With the exact expressions obtained above, we can prove rigorously the
limiting behavior of the pressure when $L\to\infty$ and $L\to 0$
discussed in section~\ref{sec:like-charge-attraction}.  

Let us consider first the case $N=2p+1$ odd. In the canonical ensemble,
the behavior of the pressure $P_c$ when $L\to \infty$, is obtained
from the term $j=0$ of (\ref{eq:ZexactNimpar}), confirming the
prediction~(\ref{eq:pLinfty}) of
section~\ref{sec:like-charge-attraction}. Furthermore, we realize that
the next to next to leading order correction is exponentially small
\begin{equation}
  P_c = -\frac{1}{4} + \frac{1}{L-2\,
    \frac{p}{p+1}} -2 \left(\frac{3p}{p+2}\right)^2 e^{-2L} 
  \left(1+O(L^{-1})\right) + O\left(e^{-6 L}\right)
  \,.
\end{equation}
In contrast, when $N=2p$, the pressure tends to 0 exponentially fast
when $L\to\infty$
\begin{equation}
  P_c = \frac{ 4p^2 e^{-L}}{(p+1)^2}
  \left(L+ \frac{2p+1}{p(p+1)}-\frac{3}{2}\right)
  +O\left( e^{-2L} \right)
  \,.
\end{equation}

The behavior of the pressure is different in the isobaric ensemble. 
Consider again first the case $N=2p+1$. From (\ref{eq:Lave}), we
already know that when $P=-1/4$, $\langle L \rangle
\to\infty$. Denoting $s=(4P+1)/2$, one can expand
(\ref{eq:Lave}) for small $s$ and invert the relation to obtain $P$ as
a function of $\langle L\rangle$ when $\langle L\rangle\to\infty$. For
instance, to order $O(s)$, Eq.~(\ref{eq:Lave}) is
\begin{equation}
  \langle L \rangle=
  \frac{4}{s}+\frac{2p}{p+1}- s S(p) + o(s)
  \,,
\end{equation}
where 
\begin{equation}
  S(p)=\sum_{k=1}^{p} \frac{1}{[k(k+1)]^2}
  =2{\cal H}^{(2)}_p - \frac{p(3p+4)}{(p+1)^2}
  \,,
\end{equation}
with ${\cal H}_{p}^{(r)} = \sum_{k=1}^{p} k^{-r} $ the harmonic
numbers. Inverting that relation, up to order $O(\langle L
\rangle^{-3})$, gives
\begin{equation}
  P=-\frac{1}{4} + 
  \frac{2}{\Lave-2\,
    \frac{p}{p+1}}
  - \frac{8S(p)}{(\Lave-2\,
    \frac{p}{p+1})^3}+o\left(\frac{1}{(\Lave-2\,
    \frac{p}{p+1})^{3}}\right)
  \,.
\end{equation}
Notice a factor 2 of difference in the next to leading order
correction (the $O(\Lave^{-1})$ term) in the pressure in the isobaric
ensemble and the canonical ensemble. Furthermore, in the isobaric
ensemble the next to next to leading order corrections are algebraic
and not exponential as in the canonical ensemble.

For $N=2p$, the behavior of the pressure, in the isobaric ensemble,
when $\Lave\to\infty$, is
\begin{equation}
  P=\frac{1}{\Lave - 2 {\cal H}_p^{(2)}}
  - \frac{2 {\cal H}_p^{(4)}}{[\Lave -2 {\cal H}_p^{(2)}]^3}
  +O\left(\Lave^{-4}\right)
  \,.
\end{equation}
Notice again the different behavior with respect to the
canonical ensemble. Here in the isobaric ensemble, the pressure vanishes
as $1/\Lave$, whereas in the canonical ensemble it vanishes
exponentially fast, as $e^{-L}$.

Let us study the other limiting behavior of the pressure, for small
separations $L$.  Let us focus on the case $N=2p+1$ first. It is not
completely straightforward to obtain the behavior of the pressure in
the canonical ensemble when $L\to 0$ directly from
expression~(\ref{eq:PexactNimpar}). Rather, it is better to return to
(\ref{eq:Zc_convol}), and notice that if $L\to 0$, then the
convolution product $f_j*f_j$ behaves as
\begin{equation}
  f_j*f_j(x) = x H(x) + O(x^2)
\end{equation}
which is independent of $j$. Then, 
\begin{equation}
 \left(
  \mathop\Asterisk_{j=1}^{p+1} f_j * f_j \right) (x) =
  \frac{x^{2p+1}}{(2p+1)!}
  + O(x^{2p+2})
\end{equation}
and
\begin{equation}
Z_c(2p+1,L)= \frac{L^{N}}{N!} + O(L^{N+1})
\,.
\end{equation}
We deduce that the pressure behaves as
\begin{equation}
  \label{eq:PcanoL0}
  P_c\sim \frac{N}{L} 
  \quad\text{when\ }L\to 0
  \,,
\end{equation}
a result already noticed in \cite{DHNP09}. Eq.~(\ref{eq:PcanoL0}) also
holds when $N=2p$.

In the isobaric ensemble, when $N=2p+1$, if $\Lave\to 0$, then,
necessarily, $s=(4P+1)/2\to\infty$ in (\ref{eq:Lave}). Expanding that
equation to order $O(s^{-2})$, one obtains
\begin{equation}
  \label{eq:PisoL0}
  P = \frac{N+1}{\Lave} 
  - \frac{N(N+2)}{12}
  +O(\Lave)
  \quad\text{when\ }\Lave\to 0
  \,.
\end{equation}
This result also holds true for $N=2p$. Notice again the difference
between the canonical (\ref{eq:PcanoL0}) and isobaric ensemble
(\ref{eq:PisoL0}), where the leading term changes from $N/L$ to
$(N+1)/L$.

When $N=2p+1$ is odd, the pressure changes of sign when $L$
varies. It is positive for $L\to 0$ and negative for $L\to \infty$. We
already obtained an approximation of the value $L^*$ of $L$ when this
occurs in the canonical ensemble, see (\ref{eq:Lstar_cano}), up to
exponentially small corrections. In the isobaric ensemble, one just
has to put $P=0$ in (\ref{eq:Lave}) to obtain the exact value
\begin{equation}
  \label{eq:Lstar_isop}
  \langle L^* \rangle =8
  \left( 
    1+ \sum_{k=1}^{p} \frac{1}{(2k+1)^2}
  \right)
  =  \pi^2-2\psi'(p+3/2)
  \,.
\end{equation}
For this quantity, the predictions from the canonical ensemble
(\ref{eq:Lstar_cano}) and the isobaric ensemble (\ref{eq:Lstar_isop})
are again different.

\begin{figure}
\begin{center}
  \includegraphics[width=7cm]{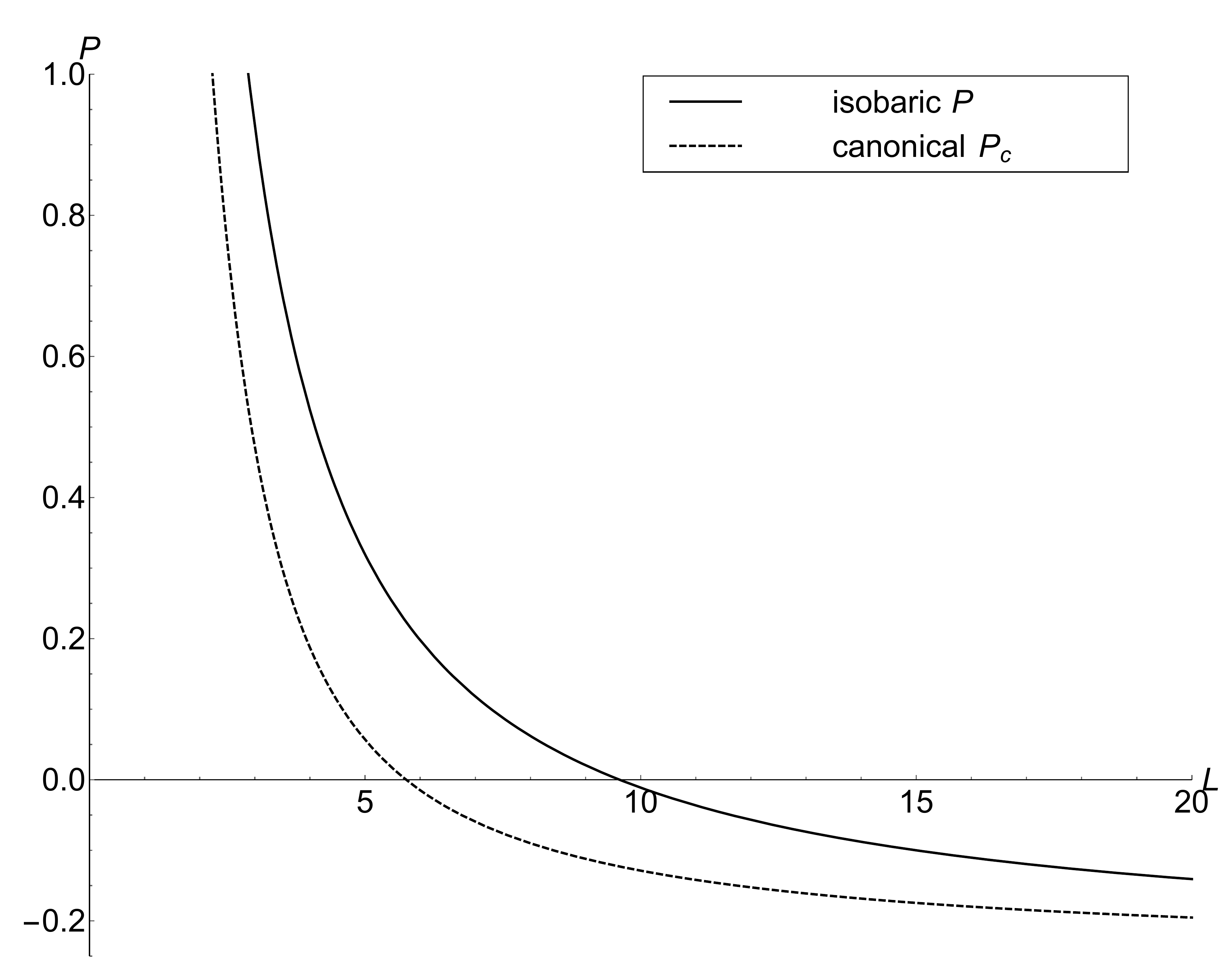}
  \caption{
    \label{fig:PvsL-iso-cano}
   The pressure $P$ as a function of the separation $L$, for
   $N=15$. The top continuous line represents the result from the
   isobaric ensemble, and the dotted bottom line those from the
   canonical ensemble. }
\end{center}
\end{figure}

\begin{figure}
\begin{center}
  \includegraphics[width=7cm]{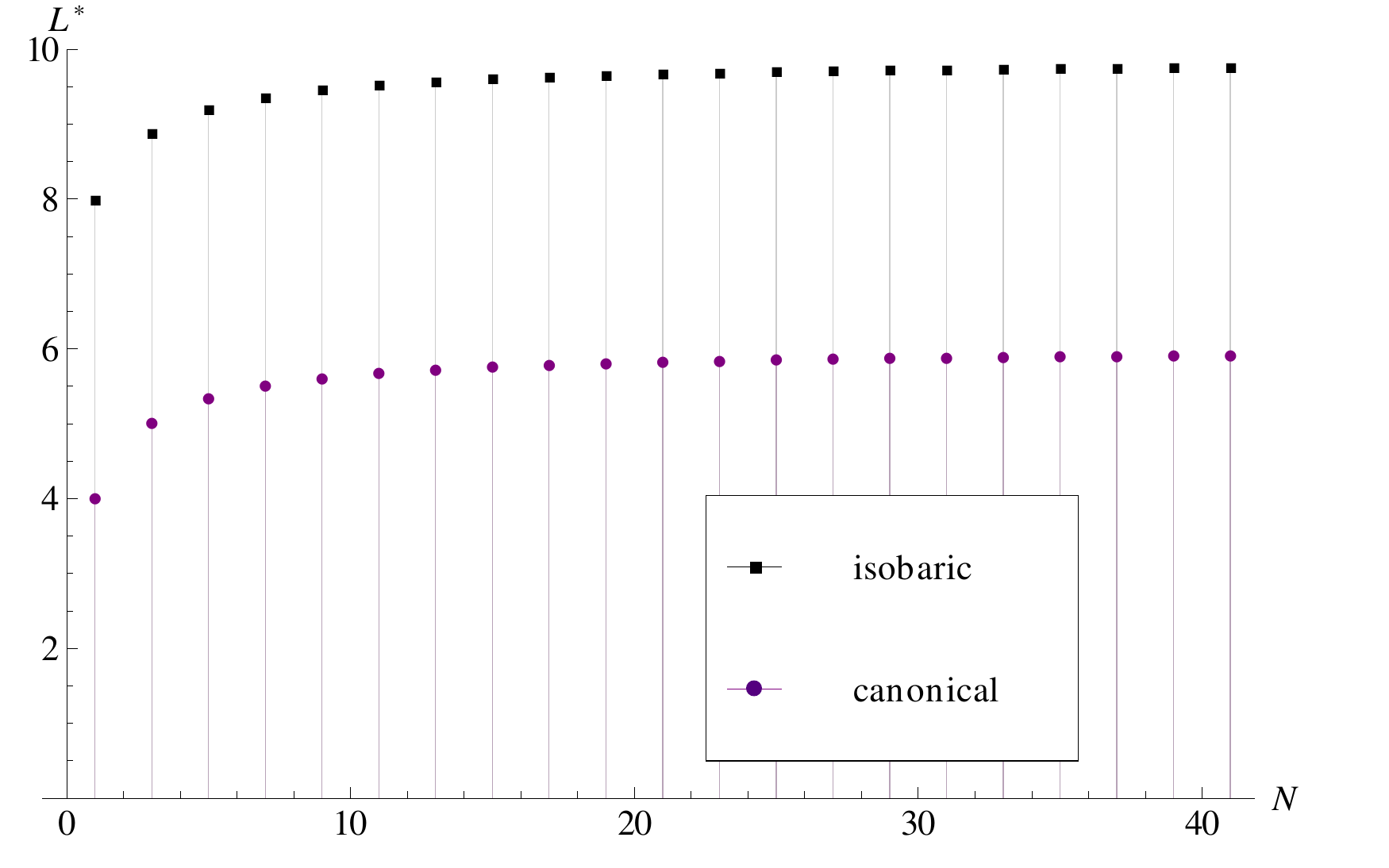}
  \caption{
    \label{fig:LstarvsN}
    The value of the separation $L^*$ for which the pressure vanishes
    and changes sign as a function of $N$ for $N$ odd. The 
    filled squares represent the results from the isobaric ensemble,
    and the filled disks, their canonical counterpart.}
\end{center}
\end{figure}

Figure~\ref{fig:PvsL-iso-cano} shows the pressure as a function of the
separation, for $N=15$, in the isobaric ensemble and the canonical
ensemble. Notice that the pressure from the canonical ensemble is
smaller that the one in the isobaric ensemble for the same separation.
%for a given value of $L$. 
Figure~\ref{fig:LstarvsN} shows the value of
$L^*$ for which the pressure changes of sign as a function of $N$,
when $N$ is odd, in both ensembles. Notice again that in the canonical
ensemble, the change of sign of the pressure occurs for smaller
values $L^*$ of the separation than in the isobaric ensemble.

%%%%%%%%%%%%%%%%%%%%%%%%%%%%%%%%%%%%%%%%%%%%%%%%%%%%%%%%%%%%%%%%%%%%%%%%%%%%%%

\section{Screening of two unequal charges}
\label{sec:different-charges}

In this section we consider a generalization of the previous model,
where the two charges located at $x=0$ and at $x=L$ are $q_1$ and
$q_2$, respectively, which can be eventually different. The overall
system should be neutral, therefore $q_1+q_2=-Ne$, $e$ being
charge of one counterion. It is convenient to introduce the notation
$Q_1$ and $Q_2$ such that $q_1=-eQ_1$ and $q_2=-eQ_2$. The
electroneutrality relation is $Q_1+Q_2=N$. The charge asymmetry can be
characterized by the quantity $a =Q_1-Q_2$, which allows to write
$Q_1=(N+a)/2$ and $Q_2=(N-a)/2$. The potential energy of the system is
now
\begin{equation}
  \label{eq:pot-a}
  U(N,L,Q_1,Q_2)=-\sum_{1\leq i < j \leq N} |{x}_i-{x}_j| +  
   a \sum_{i=1}^{N}  {x}_i
  +   (Q_2)^2 {L}
  \,.
\end{equation}
The overall effect of the charge asymmetry is to introduce a global
electric field proportional to $a$ (the term in $\sum_i x_i$).  

\subsection{Isobaric ensemble}
\label{sec:isobaric-exact-a}

Adapting the ideas of section~\ref{sec:isobaric-exact} to the
present case, we can obtain the isobaric partition function. Once again, the
results differ depending on the parity of the number of
counterions $N$. For $N=2p+1$ odd,
\begin{equation}
\label{eq:ZP-Nodd-a}
Z_P(2p+1,P,Q_1,Q_2)=\prod_{k=0}^{p}\frac{1}{\left[(k+\frac{1-a}{2})^2+ P\right]
  \left[(k+\frac{1+a}{2})^2+ P\right]}
\end{equation}
while for $N=2p$ even, 
\begin{equation}
\label{eq:ZP-Neven-a}
Z_P(2p,P,Q_1,Q_2)=\frac{1}{\left(\frac{a}{2}\right)^2+P}
\prod_{k=1}^{p}\frac{1}{\left[(k-\frac{a}{2})^2+ P\right]
  \left[(k+\frac{a}{2})^2+P\right]}
\,.
\end{equation}
The above formulas highlight the difference between the two cases,
depending on the parity of $N$. However both formulas can be
summarized in a single one as
\begin{align}
  \label{eq:ZP-a}
  Z_P(N,P,Q_1,Q_2)&=\prod_{\ell=0}^{N}\frac{1}{(\ell-\frac{N-|a|}{2})^2+P}
  =\prod_{l=0}^{N}\frac{1}{(\ell-Q_{<})^2+P}
  \nonumber\\
  &=\prod_{y\in\{-Q_<,-Q_{<}+1,\ldots, Q_{>}-1,Q_{>}\}}\frac{1}{y^2+P}
  \,.
\end{align}
where we defined
\begin{equation}
  \label{eq:Qinf}
  Q_{<}=\frac{N-|a|}{2}=\min\left(Q_1,Q_2\right)
  \quad\text{and}\quad
  Q_{>}=\frac{N+|a|}{2}=\max\left(Q_1,Q_2\right)\,.
\end{equation}
Taking the derivative of~(\ref{eq:ZP-a}) with respect to $P$, we
obtain the relation between the average length $\Lave$ of the system
and the pressure $P$ in the isobaric ensemble
\begin{equation}
  \label{eq:Lave-a}
  \Lave= \sum_{\ell=0}^{N} \frac{1}{(\ell-Q_{<})^2+P}
  \,.
\end{equation}

If $a\not\in\mathbb{Z}$ is not an integer ($q_1$ and $q_2$ are not
integer multiples of $-e/2$), or $|a|> N$ ($q_1$ and $q_2$ have
opposite signs), then $Z_P$ has simple poles. But when
$a\in\mathbb{Z}$ is an integer and $|a| \leq N$, the partition
function $Z_P$ turns out to have some double poles. This corresponds
to the case when $2Q_1$ and $2Q_2$ are both positive integers. In that
case it is best to reorder the products in~(\ref{eq:ZP-a}) to make
those double poles more apparent. The result depends on the
parity of $2Q_1$ and $2Q_2$ (both have the same parity). If $2Q_1$ and
$2Q_2$ are odd, then $Q_1$ and $Q_2$ are half integers:
$Q_1=\floor{Q_1}+\frac{1}{2}$ and $Q_2=\floor{Q_2}+\frac{1}{2}$. The
notation $\floor{x}$ denotes the floor function of $x$ (largest
integer less or equal than $x$). The isobaric partition
function~(\ref{eq:ZP-a}) becomes
\begin{equation}
  \label{eq:ZP-Na-odd}
  Z_P(N,P,Q_1,Q_2)=
  \prod_{\ell=0}^{\floor{Q_{<}}}\frac{1}{\left[(\ell+\frac{1}{2})^2+P\right]^2}
  \prod_{\ell=\floor{Q_{<}}+1}^{\floor{Q_{>}}}\frac{1}{(\ell+\frac{1}{2})^2+P}
  \,,
\end{equation}
and the corresponding equation of state is
\begin{equation}
  \label{eq:Lave-Na-odd}
  \Lave=
  \sum_{\ell=0}^{\floor{Q_{<}}}\frac{2}{(\ell+\frac{1}{2})^2+P}+
  \sum_{\ell=\floor{Q_{<}}+1}^{\floor{Q_{>}}}\frac{1}{(\ell+\frac{1}{2})^2+P}
  \,.
\end{equation}
When $Q_1$ and $Q_2$ are positive integers, these expressions become
\begin{equation}
  \label{eq:ZP-Na-even}
  Z_P(N,P,Q_1,Q_2)=\frac{1}{P}
  \prod_{\ell=1}^{Q_{<}}\frac{1}{\left(\ell^2+P\right)^2}
  \prod_{\ell=Q_{<}+1}^{Q_{>}}\frac{1}{\ell^2+P}
  \,,
\end{equation}
and
\begin{equation}
  \label{eq:Lave-Na-even}
  \Lave=\frac{1}{P}+
  \sum_{\ell=1}^{Q_{<}}\frac{2}{\ell^2+P}
  +\sum_{\ell=Q_{<}+1}^{Q_{>}}\frac{1}{\ell^2+P}
  \,.
\end{equation}

\subsection{Canonical ensemble: the partition function}

To compute the canonical partition function, one has to perform the
inverse Laplace transform of the expressions obtained in the last
section. From the above discussion, it is clear that the results will
have a different analytical structure depending on whether the
isobaric partition function has simple or double poles, that is,
depending on whether $a$ is an integer or not. If $a$ is not an
integer, or $|a|>N$, all poles of $Z_P$ are simple poles, and
we obtain from~(\ref{eq:ZP-a}):
\begin{align}
  Z_c(N,{L},Q_1,Q_2)&= 
  \sum_{j=0}^{N} (-1)^{j}e^{-(j-\frac{N-|a|}{2})^2 L}
  \frac{(2j-N+|a|)\,\Gamma(j-N+|a|)}{j!\,(N-j)!\,\Gamma(j+|a|+1)}
  \nonumber\\
 &= \sum_{j=0}^{N} (-1)^{j}e^{-(j-Q_{<})^2 L}
  \frac{2(j-Q_{<})\,\Gamma(j-2Q_{<})}{j!\,(N-j)!\,\Gamma(j+|Q_1-Q_2|+1)}
  \,.
  \label{eq:Zc-a}
\end{align}
This formula is valid whenever $2Q_1$ and $2Q_2$ are not integers, or
if $Q_1$ and $Q_2$ have opposite signs ($Q_{<}<0$ and $Q_{>}>0$).

If $a$ is an integer, with $|a| \leq N$, then
using~(\ref{eq:ZP-Na-odd}), we obtain, when $Q_1$ and $Q_2$ are half
integers,
\begin{align}
  \label{eq:Zc-Na-odd}
  Z_c(N,{L},Q_1,Q_2)&= 
  \sum_{j=0}^{\floor{Q_{<}}}
  \frac{(2j+1)^2\,e^{-(j+\frac{1}{2})^2 {L}}}
  {(\floor{Q_1}+1+j)!(\floor{Q_1}-j)!
    (\floor{Q_2}+1+j)!(\floor{Q_2}-j)!}
  \nonumber\\
  &\hspace{-2.9cm}
  \times
  \left[
    L
    -\frac{1}{2j+1}
    \left(
      \psi( \floor{Q_1}-j)
      -\psi(\floor{Q_1}+1+j)+
      \psi(\floor{Q_2}-j)
      -\psi(\floor{Q_2}+1+j)
      +\frac{2}{2j+1}
    \right)
  \right]
  \nonumber\\
  &+
  \sum_{j=\floor{Q_{<}}+1}^{\floor{Q_{>}}}
  \frac{e^{-(j+\frac{1}{2})^2{L}/4}
    (j-\floor{Q_{<}}-1)!\,(2j+1)(-1)^{j-\floor{Q_{<}}-1}
    }{(\floor{Q_{<}}+1+j)!(\floor{Q_{>}}+1+j)!(\floor{Q_{>}}-j)!}
  \,,  \nonumber\\
\end{align}
and when $Q_1$ and $Q_2$ are integers,
\begin{align}
  \label{eq:Zc-Na-even}
  Z_c(N,{L},Q_1,Q_2)&= 
  \sum_{j=1}^{Q_{<}}
  \frac{-(2j)^2\,e^{-j^2 {L}}}
  {(Q_1+j)!(Q_1-j)!(Q_2+j)!(Q_2-j)!}
  \nonumber\\
  &\hspace{-2.5cm}
  \times
  \left[
    L
    -\frac{1}{2j}
    \left(
      \psi(Q_1+1-j)
      -\psi(Q_1+1+j)+
      \psi(Q_2+1-j)
      -\psi(Q_2+1+j)
      +\frac{1}{j}
    \right)
  \right]
  \nonumber\\
  &+
  \sum_{j=Q_{<}+1}^{Q_{>}}
  \frac{e^{-j^2{L}}
    (j-Q_{<}-1)!\,(2j)(-1)^{j-Q_{<}}
  }{(Q_{<}+j)!(Q_{>}+j)!(Q_{>}-j)!}
  + 
   \frac{1}{\left(Q_1!Q_2!\right)^2}
  \,.  
\end{align}
The two previous results~(\ref{eq:Zc-Na-odd})
and~(\ref{eq:Zc-Na-even}) show the different analytical structure of
the two cases which depend on the parity of $2Q_1$ and $2Q_2$, in
particular the existence of a term independent of $L$ in the case
$2Q_1$ and $2Q_2$ even, and the form of the argument of the
exponentials $e^{-j^2 L}$ (for $2Q_1$ even), as opposed to
$e^{-(j+\frac{1}{2})^2 L}$ (for $2Q_1$ odd). However, both
results~(\ref{eq:Zc-Na-odd}) and~(\ref{eq:Zc-Na-even}) can be
subsumed in a single formula as follows. Let us define
\begin{align}
  \label{eq:ALj}
  A_j(N,{L},Q_1,Q_2)&=\frac{(2(Q_{<}-j))^2 (-1)^{2Q_{>}+1}}
  {(2Q_{<}-j)!j!(N-j)!(|Q_1-Q_2|+j)!}
  \\
  &\quad \times 
  \left[ L -  
    \frac{
      \psi({\scriptstyle j+1})-\psi({\scriptstyle 2Q_{<}-j+1})
      +\psi({\scriptstyle j+|Q_1-Q_2|+1})-\psi({\scriptstyle N-j+1})
      +\scriptstyle \frac{1}{Q_{<}-j}
    }{2(Q_{<}-j)}
  \right]
  \,,
  \nonumber
\end{align}
for $j\neq Q_{<}$, and, when $Q_{<}$ is an integer, define
\begin{equation}
  \label{eq:ALj-part}
  A_{Q_{<}}(N,{L},Q_1,Q_2)=
  \frac{1}{\left(Q_1!Q_2!\right)^2}
  \,.
\end{equation}
Also, let
\begin{equation}
  \label{eq:Dj}
  D_j(N,Q_1,Q_2)
  =\frac{j!\,2(j+Q_{<}+1)(-1)^{j+2Q_{>}+1}}{(2Q_{<}+j+1)!(N+j+1)!(|Q_1-Q_2|-j-1)!}
  \,.
\end{equation}
Then, both results~(\ref{eq:Zc-Na-odd}) and~(\ref{eq:Zc-Na-even}) are
equivalent to
\begin{equation}
  Z_c(N,{L},Q_1,Q_2)=
    \sum_{j=0}^{\floor{Q_{<}}}
    A_j(N,L,a)\, e^{-(Q_{<}-j)^2 L}
    +\sum_{j=0}^{|Q_1-Q_2|-1} D_j(N,a)\, e^{-(j+Q_{<}+1)^2 {L}}
    \,.
  \label{eq:Zc-a-general}
\end{equation}

\subsection{Canonical ensemble: asymptotic behavior of the pressure}

For small separations $L$, the results~(\ref{eq:PcanoL0}), $P_c\sim
N/L$ (canonical) and~(\ref{eq:PisoL0}), $P\sim (N+1)/\Lave$
(isobaric), still hold independently of the charge asymmetry
$a$. Thus, the effective interaction is always repulsive at short
distance, irrespective of the charges $q_1$ and $q_2$, even in the
case where these charges are opposite. Indeed, the pressure is dominated here
by the entropy cost for confining the ions in a narrow domain.

The behavior for large separations $L$ will depend on whether the
charges $q_1$ and $q_2$ are multiples of $e$ or not, and their
relative signs. There are four cases to consider.

\textbf{Opposite charges.} First, suppose that $q_1 q_2 < 0$, the
charges at the edges have opposite signs. This corresponds to the case
$|a|> N$, and the canonical partition function is obtained with
Eq.~(\ref{eq:Zc-a}). From that expression, we deduce that for $L$
large, the leading order is given by the term $j=0$ of that
sum. Therefore, the effective force is attractive and given by
\begin{equation}
  \label{eq:PcQ1Q2neg}
  P_c\sim - (Q_{<})^2 \,,\qquad L\to\infty\,,
  \,
\end{equation}
where here $Q_{<}=(N-|a|)/2<0$ corresponds to the charge of the edge
particle which has the same sign as the small ions. This result can
actually be obtained by simple arguments. The small ions will be
repelled by the particle with charge corresponding to $Q_{<}$ and
attracted to the other edge where there is a particle with charge
$-eQ_{>}$ with $Q_{>}=(N+|a|)/2>0$. By electroneutrality, the charge
of the compound object formed by the small ions and $-eQ_{>}$ will be
$eQ_{<}$. The effective force between this object and the other
opposite charge $-eQ_{<}$ is repulsive, equal to $-(eQ_{<})^2$, thus
recovering~(\ref{eq:PcQ1Q2neg}). Application of the contact theorem 
of course yields the same result, since the density of counterions 
vanishes at contact with $Q_{<}$ (a similar effect was reported in 
\cite{PaTr11,SaTr14}).

\textbf{Like-charges that are not integer multiples of $-e$.}  To
discuss this situation, we keep in mind that $Q_1>0$ and $Q_2>0$ are not
integers. The small ions of charge $e$ will be divided into two parts
that will try to screen the charges $q_1$ and $q_2$. A number
$\floor{Q_1}$ of counterions will partially screen $q_1$ and
$\floor{Q_2}$ ions will partially screen the other charge $q_2$.  Each
edge, with its screening cloud of counterions, will have a charge $-e
(Q_1-\floor{Q_1})=-e \{ Q_1 \}$ and $-e (Q_2-\floor{Q_2})=-e \{ Q_2
\}$ respectively, where $\{ x \}:=x-\floor{x}$ denotes the fractional
part of $x$. However, since $Q_1$ and $Q_2$ are not integers, we have
$\floor{Q_1}+\floor{Q_2}=N-1$: there is still one counterion to take
into consideration. This counterion feels the electric field created
by the charge difference $-e (\{Q_1\}-\{Q_2\})$, therefore it will be
attracted to the edge which has the largest remaining charge (in the
sense of the largest between $\{Q_1\}$ and $\{ Q_2 \}$). To fix the ideas
suppose $\{Q_1\}>\{Q_2\}$. The remaining ion will become part of the
screening cloud of $q_1$, and the charge of that compound object will
be $-e(\{Q_1\}-1)$. Then the effective force between the two edges
will be $e^2 (\{Q_1\}-1) \{Q_2\} = -e^2 \{Q_2\}^2$, the last equality
coming from the fact that $\{Q_1\}+\{Q_2\}=1$. Summarizing, in general
we expect an attractive force at large separations given by
\begin{equation}
  \label{eq:PcQnotint}
  P_c\sim -\left(\min\left(
      \{ Q_1 \}, \{ Q_2 \}
      \right)\right)^2\,, \qquad L\to \infty\,.
\end{equation}
This can be verified by identifying the largest argument of the
exponentials in the canonical partition function~(\ref{eq:Zc-a}) or,
equivalently, the largest pole of the isobaric partition
function~(\ref{eq:ZP-a}). The poles of the isobaric partition function
are $-(\ell - Q_{<})^2$, with $\ell$ going from 0 to $N$. Then, one
can notice that $\ell - Q_{<}$ varies from $-Q_{<}<0$ up to $Q_{>}>0$
by integer steps of 1. From this one-dimensional array of points, we
are interested in the one that is the closest to 0. That is precisely
$\min\left( \{ Q_1 \}, \{ Q_2 \} \right)$, in agreement
with~(\ref{eq:PcQnotint}). One can also notice from~(\ref{eq:Zc-a})
that in the canonical ensemble, the next to leading order correction
to~(\ref{eq:PcQnotint}) is exponentially small of order
$O(e^{-|\{ Q_1 \} - \{ Q_2 \} | L})$.

\textbf{Like-charges that are half-integer multiples of $-e$.}
A degenerate case of the previous situation is when $Q_1$ and $Q_2$
are half-integers, that is $\{Q_1\}=\{Q_2\}=\frac{1}{2}$. In this case
the canonical partition function is given by~(\ref{eq:Zc-Na-odd})
instead of~(\ref{eq:Zc-a}). The leading order is still given
by~(\ref{eq:PcQnotint}), specifically $P_c\sim -1/4$. But the
correction to leading order is not exponentially small, it can be read
from the term $j=0$ of~(\ref{eq:Zc-Na-odd})
\begin{equation}
  \label{eq:Pc-half-int}
  P_c=-\frac{1}{4}+\frac{1}{L-L_1-L_2 } + O(e^{-2L})
  \,,
\end{equation}
with
\begin{equation}
  \label{eq:L1L2}
  L_m=1-\psi(Q_m+\frac{1}{2}+1)+\psi(Q_m+\frac{1}{2})=
  \frac{Q_m-\frac{1}{2}}{Q_m+\frac{1}{2}}=
  \frac{\floor{Q_m}}{\floor{Q_m}+1}
  \,, \quad m=1,\, 2\,.
\end{equation}
We find here the generalization of the charge-symmetric case
($Q_1=Q_2=p+\frac{1}{2}$) discussed in
section~\ref{sec:equal-charges}. Each charge $q_1$ and $q_2$ is
screened by $\floor{Q_1}$ and $\floor{Q_2}$ ions. The remaining
counterion is free to roam in a region of size $L-L_1-L_2$, and with
zero electric field. This ion contributes to the pressure with a term
$\frac{1}{L-L_1-L_2}$. Here $L_1=\langle
x_{\floor{Q_1}}\rangle_{\infty}$ is the size of the screening layer of
$\floor{Q_1}$ counterions formed around $q_1$ and
$L_2=\lim_{L\to\infty} (L-\langle x_{N+1-\floor{Q_2}} \rangle)$ the
size of the layer of $\floor{Q_2}$ counterions formed around $q_2$
(compare~(\ref{eq:L1L2}) to (\ref{eq:xp}), when
$\floor{Q_1}=\floor{Q_2}=p$).

\textbf{Like-charges that are natural integer multiples of
  $-e$.}  In this case, the screening is not frustrated as in all the
previous situations. Simply $Q_1$ counterions will screen the charge
$q_1$ forming a neutral object, and similarly around $q_2$ there will
be a screening cloud of $Q_2$ counterions. Since both objects with
their screening clouds are neutral, the effective force between them is
expected to be $P_c\to 0^{+}$. This can be verified from the expression
for the partition function applicable here,
Eq.~(\ref{eq:Zc-Na-even}). If $L\to\infty$, we have
\begin{equation}
  \label{eq:Zceven-Linfty}
  Z_c=\frac{1}{Q_1!^2 Q_2!^2} \left.- \frac{4 e^{-L}}{Q_1!^2Q_2!^2}
  \frac{Q_1}{Q_1+1} \frac{Q_2}{Q_2+1} \left[ L +\frac{1}{2} \left(
    \frac{2Q_1+1}{Q_1(Q_1+1)} -
    \frac{2Q_2+1}{Q_2(Q_2+1)}-1\right)\right]\right]
    +O(e^{-4L})\,.
\end{equation}
Therefore,
\begin{equation}
  \label{eq:Pcasym-int}
  P_c= 4e^{-L}
    \frac{Q_1}{Q_1+1}    \frac{Q_2}{Q_2+1}
    \left[ L +\frac{1}{2} \left( \frac{2Q_1+1}{Q_1(Q_1+1)}
        - \frac{2Q_2+1}{Q_2(Q_2+1)}-3\right)\right]+O(e^{-2L})
    \,.
\end{equation}

\subsection{Density profile}

With the above results, we can obtain an explicit expression for the
density profile of counterions
\begin{equation}
  \label{eq:density-def}
  n({x})=\frac{\sum_{k=1}^{N}
    \int_{x_1< \cdots <x_{k-1}<x_k=x<x_{k+1}<\cdots<x_N} e^{-U(N,L,Q_1,Q_2)} \,
    \prod_{j=1,j\neq k}^{N} dx_j
    }{Z_c(N,{L},Q_1,Q_2)}
    \,.
\end{equation}
Notice that due to the fact that each particle only feels a constant
electric field proportional to the difference between the number of
charges which are at its left and right sides, the potential energy
has the following property
\begin{equation}
  \label{eq:Upot-divide}
  U(N,L,Q_1,Q_2)=
  U(k-1,x_k,Q_1,Q_2-(N-k+1))
  +U(N-k,L-x_k,Q_1-k,Q_2)
  \,.
\end{equation}
This can be interpreted as follows. If the particle at position
${x}_k$ is fixed, the system decouples into two independent
systems, one of size ${x}_k$ with $k-1$ particles, and the other
one of size ${L}-{x}_k$ with $N-k$ particles, with the
appropriate charges at each boundary (obtained by summing the charges
at the left side and right sides of ${x_k}$ of the original
system). Then, the computation of the integrals
in~(\ref{eq:density-def}) simply yields the product of the two
partition functions of each subsystem,
\begin{equation}
  \label{eq:density}
  n({x})=\frac{\sum_{k=1}^{N}
    Z_c(k-1,{x}, Q_1,Q_2-N+k-1) Z_c(N-k,{L}-{x}, Q_1 -k , Q_2)
  }{Z_c(N,{L},Q_1,Q_2)}
  \,,
\end{equation}
where each $Z_c$ should be replaced by its appropriate corresponding
expression from (\ref{eq:Zc-a}) or~(\ref{eq:Zc-a-general}).
 
\subsubsection{Contact density and pressure}

From this expression we can verify the known relation between the
contact density at $x=0$ (or $x=L$) and the
pressure~\cite{HBL79}. Indeed, notice that
\begin{equation}
  \label{eq:n0}
  n(0)=\frac{Z_c(N-1,L,Q_1-1,Q_2)}{Z_c(N,L,Q_1,Q_2)}
  \,.
\end{equation}
On the other hand, from Eq.~(\ref{eq:Zc-a}) we can verify that
\begin{equation}
  \label{eq:dZ}
  \frac{\partial Z_c(N,L,Q_1,Q_2)}{\partial L}=Z_c(N-1,L,Q_1-1,Q_2)
  -\left(Q_1\right)^2
  \,,
\end{equation}
where this last relation was obtained by writing
$-(j-\frac{N-a}{2})^2=(N-j)(j+a)-((N+a)/2)^2$ in~(\ref{eq:Zc-a}), and
recalling that $Q_1=(N+a)/2$. Therefore, we find
\begin{equation}
  \label{eq:Pcontact}
  P_c=n(0)-(Q_1)^2=n(L)-(Q_2)^2
  \,.
\end{equation}
The last equality is obtained using the same argument on $x=L$ in $n(x)$.

\subsubsection{Asymptotic behavior of the density}

\begin{figure}
\begin{center}
  \includegraphics[width=11cm]{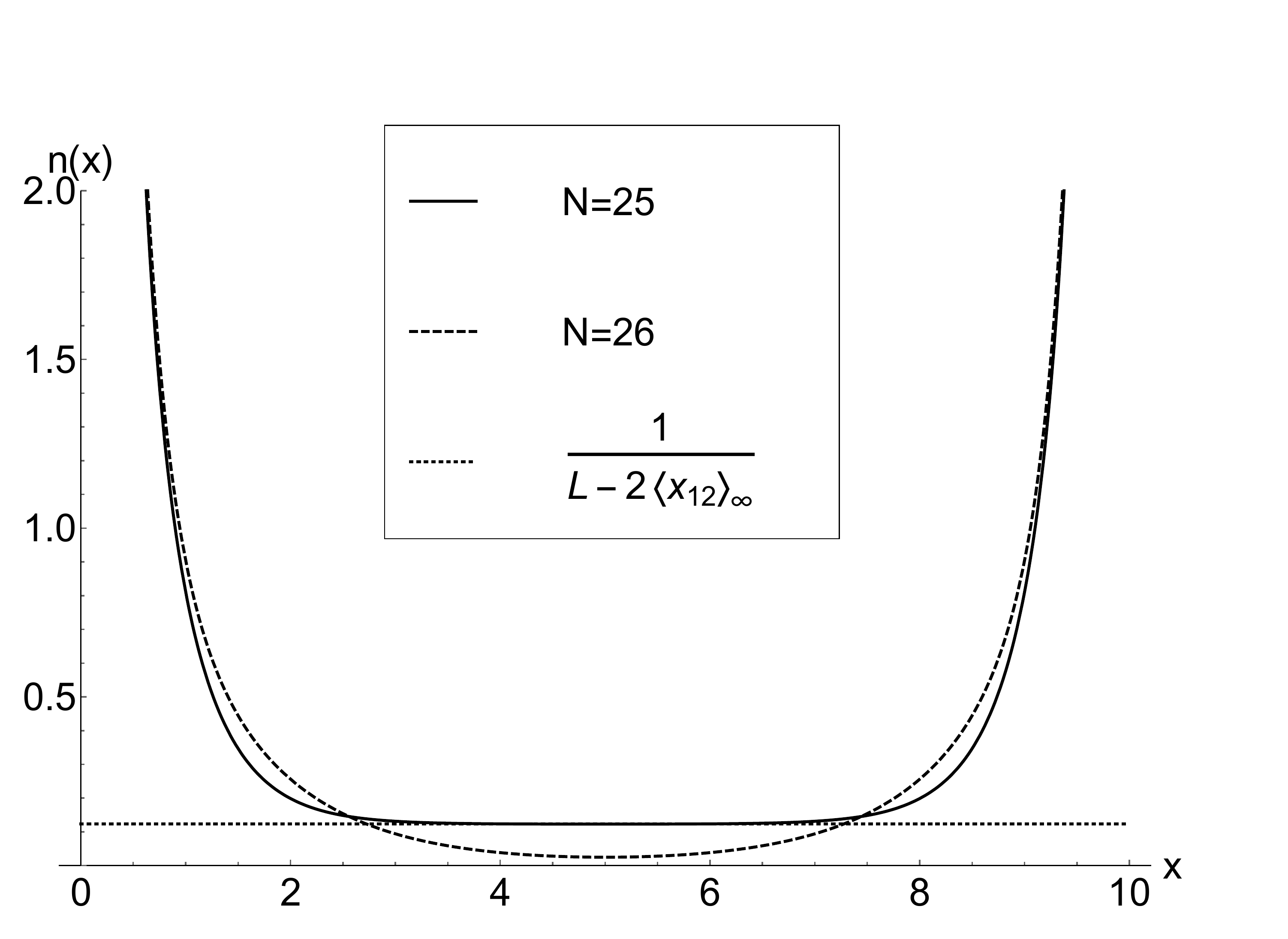}
  \caption{
    \label{fig:density}
    The density profile for $N=25$ and $N=26$ counter-ions and
    $L=10$. Notice that in the case where the number of counter-ions
    is odd, $N=25$, the density far from the edges converges to a non
    zero value $1/(L-2\langle x \rangle_{\infty})$, here
    close to $0.124$.}
\end{center}
\end{figure}

Let us consider the case $a=0$,
ie.~$Q_1=Q_2=N/2$. Figure~\ref{fig:density} shows a plot of the density
profile for $N=25$ and $N=26$. Notice that in the case $N=26$ even,
the density falls off quickly to zero far from the boundaries $x=0$ and
$x=L$. On the other hand, when $N=25$ is odd, the density does not
fall to zero, but goes to a non-vanishing value shown by the horizontal line. 
This corresponds to the density of
the free counterion, responsible for the effective attraction between
the two charges $q_1$ and $q_2$ as discussed earlier. 

To quantify this behavior, consider expression~(\ref{eq:density}) for
the density in the case $N=2p+1$, and $Q_1=Q_2=p+\frac{1}{2}$, 
\begin{equation}
  \label{eq:densitya0}
  n({x})=\frac{\sum_{k=1}^{N}
    Z_c(k-1,{x}, p+\frac{1}{2}, k-p-\frac{3}{2}) 
    Z_c(2p+1-k,{L}-{x}, p-k+\frac{1}{2}, p+\frac{1}{2} )
  }{Z_c(2p+1,{L},p+\frac{1}{2},p+\frac{1}{2})}
\,.
\end{equation}
In this sum, the partition function $Z_c(k-1,{x}, p+\frac{1}{2},
k-p-\frac{3}{2})$ corresponds to a system with charges
$-e(p+\frac{1}{2})$ and $-e(k-p-\frac{3}{2})$ at its boundaries. If
$k\leq p$, these two charges carry opposite signs, therefore,
$Z_c(k-1,{x}, p+\frac{1}{2}, k-p-\frac{3}{2})$ is given by
Eq.~(\ref{eq:Zc-a}). Then, if $1\ll x \ll L$, $Z_c(k-1,{x},
p+\frac{1}{2}, k-p-\frac{3}{2})=O(e^{-(p-k+\frac{3}{2})^2 x})$. On the
other hand, the second partition function, $Z_c(2p+1-k,{L}-{x},
p-k+\frac{1}{2}, p+\frac{1}{2} )$, corresponds to a system with
charges $-e(p-k+\frac{1}{2})$ and $-e(p+\frac{1}{2})$ at its edges. If
$k\leq p$, these two charges carry the same sign and are half integers
multiples of $e$, therefore $Z_c(2p+1-k,{L}-{x}, p-k+\frac{1}{2},
p+\frac{1}{2} )$ should be obtained by using
Eq.~(\ref{eq:Zc-Na-even}). In particular one can notice that if $1\ll
x \ll L$, then $Z_c(2p+1-k,{L}-{x}, p-k+\frac{1}{2}, p+\frac{1}{2}
)=O(e^{-(L-x)/4})$. Therefore, in the sum~(\ref{eq:densitya0}) all
terms with $k \leq p$ decay exponentially fast when $x$ is far from the
boundaries: they are of order $O\left( e^{-
    ((p-k+\frac{3}{2})^2-\frac{1}{4})x}\right)$.  The same argument
could be applied to all the terms with $k\geq p+2$, with the roles of
$Z_c(k-1,{x}, p+\frac{1}{2}, k-p-\frac{3}{2})$ and
$Z_c(2p+1-k,{L}-{x}, p-k+\frac{1}{2}, p+\frac{1}{2} )$
interchanged. Then, only one term in the sum~(\ref{eq:densitya0})
survives, it corresponds to $k=p+1$, which is precisely the index of
the position of the free counterion. In this term, both $Z_c(k-1,{x},
p+\frac{1}{2}, k-p-\frac{3}{2})$ and $Z_c(2p+1-k,{L}-{x},
p-k+\frac{1}{2}, p+\frac{1}{2} )$ with $k=p+1$, correspond to a system
with charges $-e(p+\frac{1}{2})$ and $e/2$ at its edges (notice the
opposite signs), and those partition functions should both be computed
using~(\ref{eq:Zc-a}). The leading order of these partition functions,
when $1\ll x \ll L$, is
\begin{align}
  Z_c(p,x,p+\frac{1}{2},-\frac{1}{2})\sim \frac{e^{-x/4}}{p!(p+1)!}
  &\quad\text{and}\quad
  Z_c(p,L-x,-\frac{1}{2}, p+\frac{1}{2})\sim \frac{e^{-(L-x)/4}}{p!(p+1)!}
\end{align}
while the leading order of the denominator of~(\ref{eq:densitya0}) is
\begin{equation}
  Z_c(2p+1,L,p+\frac{1}{2},p+\frac{1}{2})
  \sim \frac{e^{-L/4}}{(p!(p+1)!)^2}\left(L-2\frac{p}{p+1}\right)
  \,.
\end{equation}
This gives
\begin{equation}
  n(x)\sim \frac{1}{L-2\frac{p}{p+1}}=\frac{1}{L-2\langle x_p \rangle_{\infty}}
  \,,\quad \text{for\ } 1\ll x\ll L\,.
\end{equation}
This is the analytical confirmation of the intuitive analysis of
section \ref{sec:like-charge-attraction} where it was explained that
when $N$ is odd, there is one free ion roaming between the two charges
with an available space equal to $L-2\langle x_p \rangle_{\infty}$, as
shown in figure~\ref{fig:1Dmodel-odd}.

In the case where $N$ is even, a similar analysis shows that all terms
of the sum~(\ref{eq:density}) fall of exponentially fast when $x$ is
far from the boundaries.

\subsection{The large $N$ limit}

It is interesting to consider the limit $N\to\infty$. Due to the
electroneutrality condition $q_1+q_2+eN=0$, one needs to consider
different situations: whether $q_1$ and $q_2$ are kept finite, then
necessarily the charge of the counterions $e$ should vanish as
$1/N$. Then we notice that this is also a mean field regime. The other
possible limit is to consider that $e$ has a non vanishing finite
value, then $q_1$ and/or $q_2$ should go to infinity as $N$.

\subsubsection{Mean field limit, $N\to\infty$ and $e\to0$.}

Momentarily, it is best to return to dimensional units $\widetilde{L}$
and $\widetilde{P}$: the rescaling by $e^2$ is not appropriate here,
because $e\to0$. Consider the equation of state~(\ref{eq:Lave-a})
derived in the isobaric ensemble, which now reads
\begin{align}
  \label{eq:Lave-a-dim}
  \beta \langle \widetilde{L} \rangle
  &= \sum_{\ell=0}^{N} \frac{1}{(e\ell+q_{<})^2+\widetilde{P}}
  \sim\frac{1}{e}\int_{q_{<}}^{-q_{>}} \frac{dy}{y^2+P}
  \,,
\end{align}
where $q_{<}=-eQ_{<}$ and $q_{>}=-eQ_{>}$. Since $e\to0$, one can recognize 
a Riemann sum and replace it by an integral. This finally leads to
\begin{equation}
  \label{eq:Lave-PB}
  \beta e \langle \widetilde{L} \rangle \sqrt{\widetilde{P}} =
  \arctan\frac{q_1}{\sqrt{\widetilde{P}}}+
  \arctan\frac{q_2}{\sqrt{\widetilde{P}}}
  \,.
\end{equation}
We recover here the implicit relation between $\langle \widetilde{L}
\rangle$ and $\widetilde{P}$ from the mean field theory as described by
the Poisson--Boltzmann equation~\cite{LP99,KTNBFP08}. Indeed,
referring for instance to~\cite{KTNBFP08}, where the mean field regime
of the present problem was considered, Eq.~(\ref{eq:Lave-PB}) can be
directly obtained from a simple linear combination of Eqs.~(16) and
(17) of \cite{KTNBFP08}. Notice that the
interesting effects, such as like-charge attraction, stemming from the
discrete nature of the charges, are lost in this mean field
limit. Like-charges will always have a repulsive effective interaction
in the mean field regime \cite{lca1,lca2,lca3}. A related comment is that the asymptotic 
negative pressure reported for odd $N$ in section \ref{sec:equal-charges}, $\widetilde P = 
-q^2/N^2$, vanishes in the limiting process addressed here.

It should be noted that the present limit is also the thermodynamic
limit, since we have to remember that $e$ is of order $1/N$, therefore
in the left hand side of~(\ref{eq:Lave-PB}) $\langle\widetilde{L}\rangle$
should be of order $N$. To make this more apparent, introduce the
average distance per ion $\langle \widetilde{\ell} \rangle =
\langle\widetilde{L}\rangle/N$ (inverse of the density),
then~(\ref{eq:Lave-PB}) becomes
\begin{equation}
  \label{eq:Lave-PB-b}
  \beta (q_1+q_2) \langle \widetilde{\ell} \rangle \sqrt{\widetilde{P}} =
  \arctan\frac{q_1}{\sqrt{\widetilde{P}}}+
  \arctan\frac{q_2}{\sqrt{\widetilde{P}}}
  \,.
\end{equation}

\subsubsection{Limit $N\to\infty$ and $e$ fixed.}

In this situation, the charges at the edges $q_1$ and $q_2$ should be
of order $N$, or at least one of them. Consider the case when
both $Q_1>0$ and $Q_2>0$ are of order $N$. Then, when $N\to\infty$,
Eq.~(\ref{eq:Lave-a}) can be put in the following form
by shifting the index of the summation by $\floor{Q_{<}}$,
\begin{equation}
  \Lave =\sum_{\ell=-\infty}^{\infty} \frac{1}{(\ell-\{Q_{<}\})^2+P}
  \,.
\end{equation}
Notice that by shifting the index $\ell$ by one, we can replace
$\{Q_{<}\}$ by $\{Q_{>}\}$ if necessary. One can then write
\begin{equation}
  \label{eq:Lave-inf}
  \Lave =\sum_{\ell=-\infty}^{\infty} \frac{1}{(\ell-\min(\{Q_{1}\},\{Q_{2}\}))^2+P}
  \,.
\end{equation}
Notice that in this analysis, the limit depends on how $Q_1$ and $Q_2$
are taken to infinity, and assumes that the fractional part of them is
kept fixed as $N$ is increased.

To cover the whole range of values for $\Lave$ from 0 to $+\infty$, it
is necessary that $P$ covers the range from
$-\min(\{Q_{1}\},\{Q_{2}\})^{2}$ to $+\infty$. We recover the same
phenomenology as in the case $N$ finite, when $\Lave\to\infty$,
$P\to-\min(\{Q_{1}\},\{Q_{2}\})^{2}$. So, the pressure can become
attractive, except in the case where $Q_1$ and $Q_2$ are integers.
Eq.~(\ref{eq:Lave-inf}) can be made more explicit in two particular
cases. When $Q_1$ and $Q_2$ are integers,
\begin{equation}
  \Lave =\sum_{\ell=-\infty}^{\infty} \frac{1}{\ell^2+P}=
  \frac{\pi\coth(\pi\sqrt{P})}{\sqrt{P}}
  \,,
\end{equation}
and when $Q_1$ and $Q_2$ are half integers,
\begin{equation}
  \Lave =\sum_{\ell=-\infty}^{\infty} \frac{1}{(\ell+\frac{1}{2})^2+P}=
  \frac{\pi\tanh(\pi\sqrt{P})}{\sqrt{P}}
  \,.
\end{equation}
When $Q_1$ and $Q_2$ are not integers, the value of $\Lave$ for which
the pressure changes of sign is given by putting $P=0$
in~(\ref{eq:Lave-inf})
\begin{equation}
  \label{eq:LstarNinfty}
  \langle L^{*} \rangle=\sum_{\ell=-\infty}^{\infty}
  \frac{1}{(\ell-\min(\{Q_1\},\{Q_2\}))^2}=\psi'(\{Q_1\})+\psi'(\{Q_2\})
  \,.
\end{equation}
When $Q_1$ and $Q_2$ are half-integers this reduces to $\langle L^{*}
\rangle=\pi^{2}$.

\section{Conclusion}

We have studied a simple one-dimensional system as a model to
understand the effective interaction between charged particles that
are screened by counterions only. This model evidences the
possibility of attraction between two like-charges at large separation. 
The physical phenomenon behind this attraction
is a frustration of the screening process due to
the discrete nature of the electric charges.
More specifically, if the
two like-charges are not integers multiples of the charge of the
counterions, a perfect screening of the charges is not possible, and
there will be a ``misfit'' counterion,
responsible for the over-screening of one of the like-charges, leading
to an effective attractive force. A by-product is that in the mean-field
limit where discreteness effects are washed out, no like-charge attraction 
is possible, a well-known phenomenon.

The present model is in addition interesting from a purely theoretical
perspective, since it is exactly solvable: it is
possible to compute explicitly its partition functions (isobaric and
canonical), the pressure (effective force) and the density profile of
the counterions. Although the specific exact results and expression
for the effective force are particular to this one-dimensional model,
the physical mechanism responsible for the attraction between
like-charges could also be applicable for three dimensional
situations~\cite{MHK00}. In particular the case $N=1$ leads to an
equation of state that is equivalent to that found under strong coupling
for three dimensional planar interfaces, screened by point counter-ions
interacting through the standard $1/r$ Coulomb potential \cite{Netz01,SaTr11,Varenna}.

\textbf{Acknowledgements. } This work was supported by an ECOS
Nord/COLCIENCIAS-MEN-ICETEX action of Colombian and French
cooperation. G.~T.~acknowledges support from Fondo de Investigaciones,
Facultad de Ciencias, Universidad de los Andes, project ``Apantallamiento y 
atracci\'on de cargas similares en sistemas de Coulomb de una dimensi\'on'', 2015-2.

\begin{appendix}
\section{Two equal charges: canonical expressions}
\label{app:A}
The inverse Laplace transform can be computed with integral inversion
formula which can be evaluated using the residue theorem
\begin{equation}
  {\cal L}^{-1}\left (   \prod_{k=0}^{p}
    \frac{1}{\left[\left(k+\frac{1}{2}\right)^2+P\right]^2}
  \right) (L)
  = \sum_{j=0}^{p}
  \mathop\text{Res}_{P=-(j+\frac{1}{2})^2}
  \frac{ e^{PL}}{\prod_{k=0}^{p}\left[\left(k+\frac{1}{2}\right)^2+P\right]^2}
\end{equation}
Each residue is straightforward to compute
\begin{equation}
\mathop\text{Res}_{P=-(j+\frac{1}{2})^2} 
  \frac{e^{PL}}{ \prod_{k=0}^{p}\left[\left(k+\frac{1}{2}\right)^2+P\right]^2}
  = 
  \frac{e^{-\left(j+\frac{1}{2}\right)^2L}}{
    \prod_{k=0,k\neq j}^{p}\left[\left(k+\frac{1}{2}\right)^2-
      \left(j+\frac{1}{2}\right)^2\right]^2}
  \left( L - \sum_{l=0, l\neq j}^{p} 
    \frac{2}{\left(l+\frac{1}{2}\right)^2-\left(j+\frac{1}{2}\right)^2}
  \right)
  \,.
\end{equation}
Writing 
\begin{equation}
\frac{1}{(k+\frac{1}{2})^2-(j+\frac{1}{2})^2}=
\frac{1}{(k-j)(k+j+1)}=
\frac{1}{2j+1}\left(\frac{1}{k-j}-\frac{1}{k+j+1}\right),  
\end{equation}
the above product and sum can be simplified
\begin{equation}
  \frac{1}{\prod_{k=0,k\neq j}^{p}\left[\left(k+\frac{1}{2}\right)^2-
      \left(j+\frac{1}{2}\right)^2\right]}
  =\frac{(-1)^{j}
    (2j+1)}{(p-j)!(p+j+1)!}
  \,,
\end{equation}
and
\begin{eqnarray}
  \sum_{l=0, l\neq j}^{p} 
  \frac{1}{\left(l+\frac{1}{2}\right)^2-\left(j+\frac{1}{2}\right)^2}
  &=&
  \frac{2}{2j+1}\left(
  \frac{1}{2j+1}-\sum_{k=p-j+1}^{p+j+1} \frac{1}{k}
  \right)
  \\
  &=&\frac{2}{2j+1}\left(
  \frac{1}{2j+1}+\psi(p-j+1)-\psi(p+j+2)
  \right)
  \,.
  \nonumber
\end{eqnarray}
Gathering all results, the exact explicit result for the canonical
partition function is found in the form of Eq. (\ref{eq:ZexactNimpar}).
\end{appendix}


\begin{thebibliography}{00}

\bibitem{KeHP01}
Electrostatic Effects in Soft Matter and Biophysics, edited by P. Kekicheff,
C. Holm, and R. Podgornik,  (Kluwer Academic, Dordrecht, 2001).

\bibitem{Levin02}
Y. Levin, Rep. Prog. Phys. {\bf 65}, 1577 (2002).

\bibitem{Messina09}
R. Messina, J. Phys.: Condens. Matter {\bf 21}  113102 (2009).

\bibitem{Janco81}
B. Jancovici, Phys. Rev. Lett. {\bf 46}, 386 (1981).

\bibitem{Forrester98}
P. J. Forrester, Phys. Rep. {\bf 301}, 235 (1998).

\bibitem{Samaj03}
L. \v{S}amaj, 
J. Phys. A: Math. Gen. {\bf 36}, 5913 (2003).

\bibitem{Len61} 
A. Lenard, J. Math. Phys. \textbf{2}, 682 (1961).

\bibitem{Pra61} 
S. Prager, Adv. Chem. Phys. \textbf{4}, 201 (1961).

\bibitem{EL62} 
S. Edwards and A.~Lenard, J.~Math.~Phys.~\textbf{3},  778 (1962).

\bibitem{DHNP09} D. S. Dean, R. R. Horgan, A. Naji, R. Podgornik,
  J. Chem. Phys. \textbf{130}, 094504 (2009).
  
\bibitem{Varenna}
E. Trizac and L. \v{S}amaj, 
Proceedings of the International School of Physics Enrico Fermi {\bf 184}, 61  (2013),
edited by C. Bechinger, F. Sciortino and P. Ziherl.

\bibitem{Netz01}
R. R. Netz, Eur. Phys. J. E {\bf 5}, 557 (2001).

\bibitem{HBL79} 
D. Henderson, L. Blum and J. L. Lebowitz,
J. Electroanal. Chem.~\textbf{102}, 315 (1979).
  
\bibitem{contact2}
S. L. Carnie and D. Y .C. Chan, J. Chem. Phys. {\bf 74}, 1293 (1981).

\bibitem{contact3}
H. Wennerstr\"om, B. J\"onsson, and P. Linse,
J. Chem. Phys. {\bf 76}, 4665 (1982).

\bibitem{MaTT15}
J. P. Mallarino,   G. T\'ellez and  E. Trizac,  
Molecular Physics, (2015).

\bibitem{SaTr11}
L. \v{S}amaj and E. Trizac, 
Phys. Rev. Lett. {\bf 106}, 078301 (2011).

\bibitem{rque9}
Note that the two ions $q$, in the present one dimensional setup, play the
role of a planar confining interface, which guarantees the validity of the
contact theorem \cite{MaTT15}.

\bibitem{MHK00} R.~Messina, C.~Holm, K.~Kremer,
  Phys.~Rev.~Lett.~\textbf{85}, 872 (2000). 

\bibitem{Kim14} W. K. Kim, private communication.

\bibitem{lca1}
J. Neu, Phys. Rev. Lett. {\bf 82}, 1072 (1999).

\bibitem{lca2}
J. Sader and D. Chan, Langmuir {\bf 16}, 234 (2000).

\bibitem{lca3}
E. Trizac, Phys. Rev. E {\bf 62}, R1465 (2000).	

\bibitem{PaTr11}
F. Paillusson and E. Trizac,  
Phys. Rev. E {\bf 84}, 011407 (2011).

\bibitem{SaTr14}
L. \v{S}amaj and  E. Trizac,  
J. Stat. Phys. {\bf 156}, 932 (2014).

\bibitem{LP99} 
A. W. C. Lau and P. Pincus, Eur. Phys. J B \textbf{10},  175 (1999).

\bibitem{KTNBFP08} 
M. Kandu\v{c}, M. Trulsson, A. Naji, Y. Burak,
J. Forsman, and R. Podgornik, Phys. Rev. E \textbf{78}, 061105  (2008).




\end{thebibliography}
\end{document}